\documentclass[showpacs,preprintnumbers,twocolumn,superscriptaddress,aps,nofootinbib,reprint]{revtex4-1}
\usepackage{amssymb}
\usepackage[centertags]{amsmath}
\usepackage{txfonts}
\usepackage{epsfig}
\usepackage{bm}
\usepackage{color}
\usepackage{graphicx,graphics}
\usepackage{multirow}
\usepackage{float}
\usepackage{ulem}
\usepackage{appendix}
\usepackage[pdfstartview=FitH]{hyperref}
\hypersetup{colorlinks=true, citecolor=blue,linkcolor=blue,filecolor=black,urlcolor=blue}  % comment when submit
\allowdisplaybreaks[2]

\begin{document}

\title{Production properties of deuterons, helions and tritons via an analytical nucleon coalescence method in Pb-Pb collisions at $\sqrt{s_{NN}}=2.76$ TeV}

\author{Rui-Qin Wang}
%\email {wangrq@qfnu.edu.cn}
\affiliation{School of Physics and Physical Engineering, Qufu Normal University, Shandong 273165, China}

\author{Yan-Hao Li}
%\email{yanhaoli155@163.com}
\affiliation{School of Physics and Physical Engineering, Qufu Normal University, Shandong 273165, China}

\author{Jun Song}
%\email {songjun2011@jnxy.edu.cn}
\affiliation{School of Physical Science and Intelligent Engineering, Jining University, Shandong 273155, China}

\author{Feng-Lan Shao}
\email {shaofl@mail.sdu.edu.cn}
\affiliation{School of Physics and Physical Engineering, Qufu Normal University, Shandong 273165, China}

\begin{abstract}

We improve a nucleon coalescence model to include the coordinate-momentum correlation in nucleon joint distributions, and apply it to Pb-Pb collisions at $\sqrt{s_{NN}}=2.76$  TeV to study production properties of deuterons ($d$), helions ($^3$He) and tritons ($t$).
We give formulas of the coalescence factors $B_2$ and $B_3$, and naturally explain their behaviors as functions of the collision centrality and the transverse momentum per nucleon $p_T/A$.
We reproduce the transverse momentum spectra, averaged transverse momenta and yield rapidity densities of $d$, $^3$He and $t$, and find the system effective radius obtained in the coalescence production of light nuclei behaves similarly to Hanbury Brown-Twiss interferometry radius.
We particularly give expressions of yield ratios $d/p$, $^3$He$/d$, $t/p$, $^3$He$/p$, $d/p^{2}$, $^3$He$/p^3$, $t/^3$He and argue their nontrivial behaviors can be used to distinguish production mechanisms of light nuclei.

\end{abstract}

\maketitle

%---------------------------------------------------------------------------------------------------------------------------------------------------------------------------------------- Introduction
\section{Introduction}  \label{Introduction}

In ultra-relativistic heavy ion collisions, light nuclei such as deuterons ($d$), helions ($^3$He) and tritons ($t$) are a special group of observerables~\cite{Nagle:1994hm,Chen:2018tnh,Blum:2019suo,Bazak:2020wjn,Gutbrod:1976zzr,Aichelin:1991xy,Andronic:2017pug,Bzdak:2019pkr,Sun:2017xrx,Sun:2018jhg,Luo:2020pef,Junnarkar:2019equ,Morita:2019rph}.
They are composite clusters and their production mechanisms are still under debate so far.
As most of them are formed at the late stage of the system evolution, light nuclei are considered as sensitive probes of the fireball freeze-out properties~\cite{Nagle:1994hm,Chen:2018tnh,Blum:2019suo,Bazak:2020wjn,Gutbrod:1976zzr}. The study of light nuclei production can help understand many fundamental issues in relativistic heavy ion collision physics, e.g., the hadronization mechanism~\cite{Aichelin:1991xy}, the structure of the quantum chromodynamics phase diagram~\cite{Andronic:2017pug,Sun:2017xrx,Sun:2018jhg,Bzdak:2019pkr,Luo:2020pef} and the search for dibaryons and other molecular states~\cite{Junnarkar:2019equ,Morita:2019rph}, etc.

In recent decades, the production of light nuclei in ultra-relativistic heavy ion collisions has always attracted much attention both in experiment~\cite{STAR:2001pbk,PHENIX:2007tef,NA49:2016qvu,Albergo:2002gi,ALICE:2015wav,ALICE:2020chv,STAR:2016ydv,STAR:2020hya,Zhang:2020ewj,STAR:2019sjh,STAR:2022hbp} and in theory~\cite{Braun-Munzinger:2018hat,Oliinychenko:2020ply,Dover:1991zn,Chen:2003qj,Mrowczynski:2020ugu,Andronic:2010qu,Donigus:2022haq}.
The STAR experiment at the BNL Relativistic Heavy Ion Collider (RHIC) and the ALICE experiment at the CERN Large Hadron Collider (LHC) have collected a wealth of data on light nuclei production.
These data exhibit some fascinating features~\cite{ALICE:2015wav,ALICE:2020chv,STAR:2016ydv,STAR:2020hya,Zhang:2020ewj,STAR:2019sjh,STAR:2022hbp}.
In theory two production mechanisms, the thermal production mechanism~\cite{Mekjian:1977ei,Siemens:1979dz,Andronic:2010qu,Cleymans:2011pe,Cai:2019jtk} and the coalescence mechanism~\cite{Schwarzschild:1963zz,Sato:1981ez,Dover:1991zn,Mattiello:1995xg,Nagle:1996vp,Mattiello:1996gq,Chen:2003qj,Polleri:1997bp,Scheibl:1998tk,Sharma:2018dyb,Bazak:2018hgl}, have proved to be successful in describing light nuclei formation. 
In addition, transport scenario~\cite{Danielewicz:1991dh,Oh:2009gx,Oliinychenko:2018ugs,Oliinychenko:2020znl,Staudenmaier:2021lrg,Kireyeu:2022qmv,Coci:2023daq} is employed to study how light nuclei evolve and survive during the hadronic system evolution.

The coalescence mechanism, in which light nuclei are assumed to be produced by the coalescence of the jacent nucleons in the phase space, possesses its unique characteristics.
In order to see whether, if so, to what extent, these characteristics depend on the particular coalescence model used in obtaining these characteristics, 
we in our previous works~\cite{Zhao:2022xkz,Wang:2022hja,Wang:2020zaw} developed an analytic description for the production of different species of light nuclei in the coalescence picture with the assumption of the coordinate-momentum factorization.
The obtained analytic formulas clearly show the relationships of light nuclei with primordial nucleons and effects of different factors on light nuclei production such as the whole hadronic system scale as well as the sizes of the formed light nuclei.
In Refs.~\cite{Zhao:2022xkz,Wang:2022hja}, we applied the analytic coalescence model to Au-Au collisions at RHIC energies to successfully explain the transverse momentum spectra, yield rapidity densities, averaged transverse momenta and yield correlations of different light nuclei.
We also applied it to pp, p-Pb and Pb-Pb collisions at LHC to study the behavior of the coalescence factor $B_A$~\cite{Wang:2020zaw}, and found it can naturally explain the relatively weak $p_T$ dependence of $B_A$ in pp and p-Pb collisions.
In Pb-Pb collisions it gave qualitative growth of $B_A$ against $p_T$, but growth extent was underestimated.
It is urgently necessary to give quantitative explanations for $B_A$ and further explore intrinsic properties of light nuclei production in heavy ion collisions with such high collision energy at the LHC.

In this work, we extend the nucleon coalescence model to include the coordinate-momentum correlation originating possibly from the collective flows~\cite{Csorgo:1995bi}, the temperature gradients~\cite{Tomasik:1997eq}, etc., and apply it to Pb-Pb collisions at LHC to study the production of light nuclei. 
One main goal of this article is to bring to light the characteristics originating from the nucleon coalescence itself and to discriminate influences of different factors in heavy ion collisions with so high collision energy on light nuclei production.
We study coalescence factors ($B_2$ and $B_3$), transverse momentum ($p_T$) spectra, averaged transverse momenta ($\langle p_T\rangle$), yield rapidity densities ($dN/dy$) and yield ratios of different species of light nuclei.
We find the nucleon coalescence model including the coordinate-momentum correlation can well describe the light nuclei production in Pb-Pb collisions at $\sqrt{s_{NN}}=2.76$ TeV.
We also find the system effective radius obtained in the coalescence production of light nuclei behaves similarly to Hanbury Brown-Twiss (HBT) interferometry radius.

The paper is organized as follows. 
In Sec.~\ref{Model} we give an introduction to the nucleon coalescence model. 
In Sec.~\ref{BA} we study coalescence factors $B_2$ and $B_3$, and discuss their behaviors as functions of the collision centrality and the transverse momentum per nucleon.
In Sec.~\ref{pTspectra}, we study the $p_T$ spectra, averaged transverse momenta, yield rapidity densities and yield ratios of  $d$, $^3$He and $t$.
In Sec.~\ref{Summary} we give our summary.
 
%-------------------------------------------------------------------------------------------------------------------------------------------------------------------------------------------------------------------------------------------------------- Model
\section{The nucleon coalescence model}  \label{Model}

In this section we extend the nucleon coalescence model in our previous works~\cite{Wang:2020zaw,Zhao:2022xkz,Wang:2022hja} to include the coordinate-momentum correlation in nucleon joint distributions.
We present formulism of two nucleons coalescing into $d$ and that of three nucleons coalescing into $^3$He and $t$.
For $t$, the deduction process is the same as that of $^3$He and we do not repeat the display and only give the final formula.

%-------------------------------------------------------------------------------------------------- general formalism
We start from a hadronic system produced at the final stage of the evolution of high energy heavy ion collision and suppose light nuclei are formed via the nucleon coalescence.
The three-dimensional momentum distribution of the produced deuterons $f_{d}(\boldsymbol{p})$ and that of helions $f_{\mathrm{^{3}He}}(\boldsymbol{p})$ are
{\setlength\arraycolsep{0pt}
\begin{eqnarray}
 f_{d}(\boldsymbol{p})&=& N_{pn} \int d\boldsymbol{x}_1d\boldsymbol{x}_2 d\boldsymbol{p}_1 d\boldsymbol{p}_2  f^{(n)}_{pn}(\boldsymbol{x}_1,\boldsymbol{x}_2;\boldsymbol{p}_1,\boldsymbol{p}_2)  \nonumber  \\
   &&~~~~~~~~~~~ \times  \mathcal {R}_{d}(\boldsymbol{x}_1,\boldsymbol{x}_2;\boldsymbol{p}_1,\boldsymbol{p}_2,\boldsymbol{p}),      \label{eq:fdgeneral}   \\
 f_{\mathrm{^{3}He}}(\boldsymbol{p})&=& N_{ppn} \int d\boldsymbol{x}_1d\boldsymbol{x}_2d\boldsymbol{x}_3 d\boldsymbol{p}_1 d\boldsymbol{p}_2 d\boldsymbol{p}_3 \nonumber  \\
   &&~~~~~~~~~~~~ \times  f^{(n)}_{ppn}(\boldsymbol{x}_1,\boldsymbol{x}_2,\boldsymbol{x}_3;\boldsymbol{p}_1,\boldsymbol{p}_2,\boldsymbol{p}_3) \nonumber  \\
   &&~~~~~~~~~~~~ \times \mathcal {R}_{\mathrm{^{3}He}}(\boldsymbol{x}_1,\boldsymbol{x}_2,\boldsymbol{x}_3;\boldsymbol{p}_1,\boldsymbol{p}_2,\boldsymbol{p}_3,\boldsymbol{p}).     \label{eq:fHe3general} 
\end{eqnarray} }%
Here $f^{(n)}_{pn}(\boldsymbol{x}_1,\boldsymbol{x}_2;\boldsymbol{p}_1,\boldsymbol{p}_2)$ is the normalized joint coordinate-momentum distribution of proton-neutron pairs and $f^{(n)}_{ppn}(\boldsymbol{x}_1,\boldsymbol{x}_2,\boldsymbol{x}_3;\boldsymbol{p}_1,\boldsymbol{p}_2,\boldsymbol{p}_3)$ is that of three-nucleon clusters. 
$N_{pn}=N_{p}N_{n}$ is the number of all possible $pn$-pairs and $N_{ppn}=N_{p}(N_{p}-1)N_{n}$ is that of all possible $ppn$-clusters.
$N_{p}$ is the proton number and $N_{n}$ is the neutron number in the considered hadronic system.
$\mathcal {R}_{d}(\boldsymbol{x}_1,\boldsymbol{x}_2;\boldsymbol{p}_1,\boldsymbol{p}_2,\boldsymbol{p})$ and $\mathcal {R}_{\mathrm{^{3}He}}(\boldsymbol{x}_1,\boldsymbol{x}_2,\boldsymbol{x}_3;\boldsymbol{p}_1,\boldsymbol{p}_2,\boldsymbol{p}_3,\boldsymbol{p})$ are kernel functions.
Here and from now on we use boldface type to distinguish three-dimensional vectors.

Taking into account constraints from the momentum conservation and intrinsic quantum numbers of light nuclei, we rewrite kernel functions in the following forms as in Refs.~\cite{Wang:2020zaw,Zhao:2022xkz,Wang:2022hja,Wang:2017vsm}
{\setlength\arraycolsep{0pt}
\begin{eqnarray}
&&  \mathcal {R}_{d}(\boldsymbol{x}_1,\boldsymbol{x}_2;\boldsymbol{p}_1,\boldsymbol{p}_2,\boldsymbol{p}) = g_d \mathcal {R}_{d}^{(x,p)}(\boldsymbol{x}_1,\boldsymbol{x}_2;\boldsymbol{p}_1,\boldsymbol{p}_2)  
  \delta(\displaystyle{\sum^2_{i=1}} \boldsymbol{p}_i-\boldsymbol{p}),~~~~  \label{eq:Rd}  \\
&&  \mathcal {R}_{\mathrm{^{3}He}}(\boldsymbol{x}_1,\boldsymbol{x}_2,\boldsymbol{x}_3;\boldsymbol{p}_1,\boldsymbol{p}_2,\boldsymbol{p}_3,\boldsymbol{p}) = g_{\mathrm{^{3}He}}  \nonumber  \\
&&~~~~~~~~~~~~~~~~~~~~~~~~ \times \mathcal {R}_{\mathrm{^{3}He}}^{(x,p)}(\boldsymbol{x}_1,\boldsymbol{x}_2,\boldsymbol{x}_3;\boldsymbol{p}_1,\boldsymbol{p}_2,\boldsymbol{p}_3)   \delta(\displaystyle{\sum^3_{i=1}} \boldsymbol{p}_i-\boldsymbol{p}) ,      \label{eq:RHe3}  
\end{eqnarray} }%
where the spin degeneracy factors $g_d=3/4$ and $g_{\mathrm{^{3}He}}=1/4$.
The Dirac $\delta$ functions guarantee the momentum conservation in the coalescence process.
The remaining $\mathcal {R}_{d}^{(x,p)}(\boldsymbol{x}_1,\boldsymbol{x}_2;\boldsymbol{p}_1,\boldsymbol{p}_2)$ and
$\mathcal {R}_{\mathrm{^{3}He}}^{(x,p)}(\boldsymbol{x}_1,\boldsymbol{x}_2,\boldsymbol{x}_3;\boldsymbol{p}_1,\boldsymbol{p}_2,\boldsymbol{p}_3)$ can be solved from the Wigner transformation as adopting the wave function of a spherical harmonic oscillator as in Refs.~\cite{Chen:2003ava,Zhu:2015voa}.
They are as follows
{\setlength\arraycolsep{0pt}
\begin{eqnarray}
&&  \mathcal {R}^{(x,p)}_{d}(\boldsymbol{x}_1,\boldsymbol{x}_2;\boldsymbol{p}_1,\boldsymbol{p}_2) = 8e^{-\frac{(\boldsymbol{x}'_1-\boldsymbol{x}'_2)^2}{\sigma_d^2}}     e^{-\frac{\sigma_d^2(\boldsymbol{p}'_{1}-\boldsymbol{p}'_{2})^2}{4\hbar^2c^2}},      \label{eq:Rdxp}  \\
&&  \mathcal {R}^{(x,p)}_{\mathrm{^{3}He}}(\boldsymbol{x}_1,\boldsymbol{x}_2,\boldsymbol{x}_3;\boldsymbol{p}_1,\boldsymbol{p}_2,\boldsymbol{p}_3) =8^2e^{-\frac{(\boldsymbol{x}'_1-\boldsymbol{x}'_2)^2}{2\sigma_{\mathrm{^{3}He}}^2}} e^{-\frac{(\boldsymbol{x}'_1+\boldsymbol{x}'_2-2\boldsymbol{x}'_3)^2}{6\sigma_{\mathrm{^{3}He}}^2}}  \nonumber   \\
&& ~~~~~~~~~~~~~~~~~~~~~~~~~~~~~~~~~~~~\times e^{-\frac{\sigma_{\mathrm{^{3}He}}^2(\boldsymbol{p}'_{1}-\boldsymbol{p}'_{2})^2}{2\hbar^2c^2}} e^{-\frac{\sigma_{\mathrm{^{3}He}}^2(\boldsymbol{p}'_{1}+\boldsymbol{p}'_{2}-2\boldsymbol{p}'_{3})^2}{6\hbar^2c^2}}. \label{eq:RHe3xp}
\end{eqnarray} }%
The superscript `$'$' in the coordinate or momentum variable denotes the nucleon coordinate or momentum in the rest frame of the $pn$-pair or $ppn$-cluster.
The width parameter $\sigma_d=\sqrt{\frac{8}{3}} R_d$ and $\sigma_{\mathrm{^{3}He}}=R_{\mathrm{^{3}He}}$,
where $R_d=2.1421$ fm and $R_{\mathrm{^{3}He}}=1.9661$ fm are the root-mean-square radius of the deuteron and that of the $^3$He, respectively~\cite{Angeli:2013epw}.
The factor $\hbar c$ comes from the used GeV$\cdot$fm unit, and it is 0.197 GeV$\cdot$fm.

Substituting Eqs. (\ref{eq:Rd}-\ref{eq:RHe3xp}) into Eqs. (\ref{eq:fdgeneral}) and (\ref{eq:fHe3general}), we have
{\setlength\arraycolsep{0.2pt}
\begin{eqnarray}
&& f_{d}(\boldsymbol{p})= g_{d} N_{pn} \int d\boldsymbol{x}_1d\boldsymbol{x}_2d\boldsymbol{p}_1d\boldsymbol{p}_2 f^{(n)}_{pn}(\boldsymbol{x}_1,\boldsymbol{x}_2;\boldsymbol{p}_1,\boldsymbol{p}_2)   \nonumber   \\
&& ~~~~~~~~~~~~~~~~~~~~~~~~~~~ \times
8e^{-\frac{(\boldsymbol{x}'_1-\boldsymbol{x}'_2)^2}{\sigma_d^2}} e^{-\frac{\sigma_d^2(\boldsymbol{p}'_{1}-\boldsymbol{p}'_{2})^2}{4\hbar^2c^2}} \delta(\displaystyle{\sum^2_{i=1}} \boldsymbol{p}_i-\boldsymbol{p}),~~~~~~     \label{eq:fd}  \\
%----------------------------------------------------------------------------------------------------
&& f_{\mathrm{^{3}He}}(\boldsymbol{p}) = g_{\mathrm{^{3}He}} N_{ppn}  \int d\boldsymbol{x}_1d\boldsymbol{x}_2d\boldsymbol{x}_3d\boldsymbol{p}_1d\boldsymbol{p}_2d\boldsymbol{p}_3  \nonumber   \\
&& ~~~~~~ \times  f^{(n)}_{ppn}(\boldsymbol{x}_1,\boldsymbol{x}_2,\boldsymbol{x}_3;\boldsymbol{p}_1,\boldsymbol{p}_2,\boldsymbol{p}_3)   \delta(\displaystyle{\sum^3_{i=1}} \boldsymbol{p}_i-\boldsymbol{p})    \nonumber   \\
&& ~~~~~~ \times
8^2e^{-\frac{(\boldsymbol{x}'_1-\boldsymbol{x}'_2)^2}{2\sigma_{\mathrm{^{3}He}}^2}} e^{-\frac{(\boldsymbol{x}'_1+\boldsymbol{x}'_2-2\boldsymbol{x}'_3)^2}{6\sigma_{\mathrm{^{3}He}}^2}}
e^{-\frac{\sigma_{\mathrm{^{3}He}}^2(\boldsymbol{p}'_{1}-\boldsymbol{p}'_{2})^2}{2\hbar^2c^2}} e^{-\frac{\sigma_{\mathrm{^{3}He}}^2(\boldsymbol{p}'_{1}+\boldsymbol{p}'_{2}-2\boldsymbol{p}'_{3})^2}{6\hbar^2c^2}}.  \label{eq:fHe3}  
\end{eqnarray} }%

%---------------------------------------------------------------------------------------------------------------------  integrate dp1 dp2
Considering that the gaussian width values $2\hbar c/\sigma_d$, $\sqrt{2}\hbar c/\sigma_{\mathrm{^{3}He}}$ and $\sqrt{6}\hbar c/\sigma_{\mathrm{^{3}He}}$ in the momentum-dependent kernel functions are quite small,
we mathematically approximate the gaussian form $e^{-(\Delta \boldsymbol{p}')^2/\epsilon^2}$ as $(\sqrt{\pi} \epsilon)^3 \delta(\Delta \boldsymbol{p}')$, 
where $\epsilon$ is a small quantity.
Then we can obtain
{\setlength\arraycolsep{0.2pt}
\begin{eqnarray}
  f_{d}(\boldsymbol{p}) &=& 8 g_{d} N_{pn} \int d\boldsymbol{x}_1d\boldsymbol{x}_2d\boldsymbol{p}_1d\boldsymbol{p}_2 f^{(n)}_{pn}(\boldsymbol{x}_1,\boldsymbol{x}_2;\boldsymbol{p}_1,\boldsymbol{p}_2)  \nonumber   \\
&& \times  e^{-\frac{(\boldsymbol{x}'_1-\boldsymbol{x}'_2)^2}{\sigma_d^2}}   \left(\frac{2\hbar c\sqrt{\pi}}{\sigma_d}\right)^3
  \delta(\boldsymbol{p}'_{1}-\boldsymbol{p}'_{2})  \delta(\displaystyle{\sum^2_{i=1}} \boldsymbol{p}_i-\boldsymbol{p})                     \nonumber            \\                             
&=&8g_{d} N_{pn} \int d\boldsymbol{x}_1d\boldsymbol{x}_2d\boldsymbol{p}_1d\boldsymbol{p}_2 f^{(n)}_{pn}(\boldsymbol{x}_1,\boldsymbol{x}_2;\boldsymbol{p}_1,\boldsymbol{p}_2)  \nonumber   \\
&& \times  e^{-\frac{(\boldsymbol{x}'_1-\boldsymbol{x}'_2)^2}{\sigma_d^2}}   \left(\frac{2\hbar c\sqrt{\pi}}{\sigma_d}\right)^3
 \gamma \delta(\boldsymbol{p}_{1}-\boldsymbol{p}_{2})  \delta(\displaystyle{\sum^2_{i=1}} \boldsymbol{p}_i-\boldsymbol{p})          \nonumber   \\              
&=&8g_{d} N_{pn}  \left(\frac{\hbar c\sqrt{\pi}}{\sigma_d}\right)^3 \gamma    \nonumber  \\
 && \times  \int d\boldsymbol{x}_1d\boldsymbol{x}_2  f^{(n)}_{pn}(\boldsymbol{x}_1,\boldsymbol{x}_2;\frac{\boldsymbol{p}}{2},\frac{\boldsymbol{p}}{2})   e^{-\frac{(\boldsymbol{x}'_1-\boldsymbol{x}'_2)^2}{\sigma_d^2}},      \label{eq:fd-p}   
\end{eqnarray}	
where the Lorentz contraction factor $\gamma$ comes from the Lorentz transformation $\Delta \boldsymbol{p}'=\frac{1}{\gamma}\Delta \boldsymbol{p}$.
Similarly for $^3$He we have
{\setlength\arraycolsep{0.2pt}
\begin{eqnarray}
&& f_{\mathrm{^{3}He}}(\boldsymbol{p}) 
=8^{2} g_{\mathrm{^{3}He}}N_{ppn} \left(\frac{\hbar^{2} c^{2}\pi}{\sqrt{3}\sigma_\mathrm{^{3}He}^{2}}\right)^3 \gamma^{2} \times       \nonumber \\ 
&&  \int d\boldsymbol{x}_1d\boldsymbol{x}_2d\boldsymbol{x}_3 f^{(n)}_{ppn}(\boldsymbol{x}_1,\boldsymbol{x}_2,\boldsymbol{x}_3;\frac{\boldsymbol{p}}{3},\frac{\boldsymbol{p}}{3},\frac{\boldsymbol{p}}{3}) 
 e^{-\frac{(\boldsymbol{x}'_1-\boldsymbol{x}'_2)^2}{2\sigma_\mathrm{^{3}He}^2}}e^{-\frac{(\boldsymbol{x}'_1+\boldsymbol{x}'_2-2\boldsymbol{x}'_3)^2}{6\sigma_\mathrm{^{3}He}^2}}.  \nonumber  \\       \label{eq:fHe3-p}   
\end{eqnarray}	

%-------------------------------------------------------------------------------------------------- integrate dr  
Changing coordinate integral variables in Eq. (\ref{eq:fd-p}) to be $\boldsymbol{X}= \frac{\boldsymbol{x}_1+\boldsymbol{x}_2}{2}$ and $\boldsymbol{r}= \boldsymbol{x}_1-\boldsymbol{x}_2$, 
and those in Eq. (\ref{eq:fHe3-p}) to be $\boldsymbol{Y}= (\boldsymbol{x}_1+\boldsymbol{x}_2+\boldsymbol{x}_3)/\sqrt{3}$, $\boldsymbol{r}_1= (\boldsymbol{x}_1-\boldsymbol{x}_2)/\sqrt{2}$ and $\boldsymbol{r}_2= (\boldsymbol{x}_1+\boldsymbol{x}_2-2\boldsymbol{x}_3)/\sqrt{6}$,
we have
{\setlength\arraycolsep{0pt}
\begin{eqnarray}
&& f_{d}(\boldsymbol{p})= 8g_{d} N_{pn}  \left(\frac{\hbar c\sqrt{\pi}}{\sigma_d}\right)^3 \gamma   \int d\boldsymbol{X}d\boldsymbol{r}  f^{(n)}_{pn}(\boldsymbol{X},\boldsymbol{r};\frac{\boldsymbol{p}}{2},\frac{\boldsymbol{p}}{2})
  e^{-\frac{\boldsymbol{r}'^2}{\sigma_d^2}},~~~~~~      \label{eq:fd-Xr}  \\
%----------------------------------------------------------------------------------------------------
&& f_{\mathrm{^{3}He}}(\boldsymbol{p}) = 8^{2} g_{\mathrm{^{3}He}}N_{ppn} \left(\frac{\hbar^{2} c^{2}\pi}{\sqrt{3}\sigma_\mathrm{^{3}He}^{2}}\right)^3 \gamma^{2} \nonumber \\ 
&& ~~~~~ \times  \int d\boldsymbol{Y}d\boldsymbol{r}_1d\boldsymbol{r}_2  f^{(n)}_{ppn}(\boldsymbol{Y},\boldsymbol{r}_1,\boldsymbol{r}_2;\frac{\boldsymbol{p}}{3},\frac{\boldsymbol{p}}{3},\frac{\boldsymbol{p}}{3}) 
 e^{-\frac{\boldsymbol{r}_1'^2}{\sigma_\mathrm{^{3}He}^2}}e^{-\frac{\boldsymbol{r}_2'^2}{\sigma_\mathrm{^{3}He}^2}}.      \label{eq:fHe3-Yr}  
\end{eqnarray} }%
Considering the nucleon strong interaction and the nucleon coalescence are local, we neglect the effect of collective motion on the center of mass coordinate and assume it is factorized in nucleon joint distributions, i.e.,
{\setlength\arraycolsep{0pt}
\begin{eqnarray}
 && f^{(n)}_{pn}(\boldsymbol{X},\boldsymbol{r};\frac{\boldsymbol{p}}{2},\frac{\boldsymbol{p}}{2}) = f^{(n)}_{pn}(\boldsymbol{X}) f^{(n)}_{pn}(\boldsymbol{r};\frac{\boldsymbol{p}}{2},\frac{\boldsymbol{p}}{2}),   \label{eq:fpn-Xr}      \\
 && f^{(n)}_{ppn}(\boldsymbol{Y},\boldsymbol{r}_1,\boldsymbol{r}_2;\frac{\boldsymbol{p}}{3},\frac{\boldsymbol{p}}{3},\frac{\boldsymbol{p}}{3}) = f^{(n)}_{ppn}(\boldsymbol{Y}) 
 f^{(n)}_{ppn}(\boldsymbol{r}_1,\boldsymbol{r}_2;\frac{\boldsymbol{p}}{3},\frac{\boldsymbol{p}}{3},\frac{\boldsymbol{p}}{3}). ~~~~  \label{eq:fppn-Yr1r2}
\end{eqnarray} }%
Then we have
{\setlength\arraycolsep{0pt}
\begin{eqnarray}
&& f_{d}(\boldsymbol{p})= 8g_{d} N_{pn}  \left(\frac{\hbar c\sqrt{\pi}}{\sigma_d}\right)^3 \gamma   \int d\boldsymbol{r}  f^{(n)}_{pn}(\boldsymbol{r};\frac{\boldsymbol{p}}{2},\frac{\boldsymbol{p}}{2})   e^{-\frac{r'^2}{\sigma_d^2}},~~~~~     \label{eq:fd-r}  \\
%----------------------------------------------------------------------------------------------------
&& f_{\mathrm{^{3}He}}(\boldsymbol{p}) = 8^{2} g_{\mathrm{^{3}He}}N_{ppn} \left(\frac{\hbar^{2} c^{2}\pi}{\sqrt{3}\sigma_\mathrm{^{3}He}^{2}}\right)^3 \gamma^{2} \nonumber \\ 
&& ~~~~~~~~~~ \times  \int d\boldsymbol{r}_1d\boldsymbol{r}_2 f^{(n)}_{ppn}(\boldsymbol{r}_1,\boldsymbol{r}_2;\frac{\boldsymbol{p}}{3},\frac{\boldsymbol{p}}{3},\frac{\boldsymbol{p}}{3})  
  e^{-\frac{\boldsymbol{r}_1'^2}{\sigma_\mathrm{^{3}He}^2}}e^{-\frac{\boldsymbol{r}_2'^2}{\sigma_\mathrm{^{3}He}^2}}.        \label{eq:fHe3-r}  
\end{eqnarray} }%

We adopt the frequently-used gaussian form for the relative coordinate distribution as in such as Ref.~\cite{Mrowczynski:2016xqm}, i.e., 
{\setlength\arraycolsep{0pt}
\begin{eqnarray}
&& f^{(n)}_{pn}(\boldsymbol{r};\frac{\boldsymbol{p}}{2},\frac{\boldsymbol{p}}{2}) = \frac{1}{\left[\pi C R_f^2(\boldsymbol{p})\right]^{3/2}} e^{-\frac{\boldsymbol{r}^2}{C R_f^2(\boldsymbol{p})}} f^{(n)}_{pn}(\frac{\boldsymbol{p}}{2},\frac{\boldsymbol{p}}{2}),    \label{eq:fpn-rp}  \\
&& f^{(n)}_{ppn}(\boldsymbol{r}_1,\boldsymbol{r}_2;\frac{\boldsymbol{p}}{3},\frac{\boldsymbol{p}}{3},\frac{\boldsymbol{p}}{3}) = \frac{1}{\left[\pi^2 C_1C_2 R_f^4(\boldsymbol{p})\right]^{3/2}} e^{-\frac{\boldsymbol{r}_1^2}{C_1 R_f^2(\boldsymbol{p})}}  \nonumber  \\
&& ~~~~~~~~~~~~~~~~~~~~~~~~~~~~~~~~~~~~~ \times    e^{-\frac{\boldsymbol{r}_2^2}{C_2 R_f^2(\boldsymbol{p})}}   f^{(n)}_{ppn}(\frac{\boldsymbol{p}}{3},\frac{\boldsymbol{p}}{3},\frac{\boldsymbol{p}}{3}).  \label{eq:fppn-r1r2p}
\end{eqnarray} }%
Here $R_f(\boldsymbol{p})$ is the effective radius of the source system at the light nuclei freeze-out, and it generally depends on the momentum of the light nuclei~\cite{Kisiel:2014upa,ALICE:2015hvw,ALICE:2015tra}. 
Considering relations between $\boldsymbol{r}$, $\boldsymbol{r}_1$ and $\boldsymbol{r}_2$ with $\boldsymbol{x}_1$, $\boldsymbol{x}_2$ and $\boldsymbol{x}_3$, $C_1$ equals to $C/2$ 
and $C_2$ equals to $2C/3$.
So there is only one distribution width parameter $C$ to be determined, and it is set to be 4 the same as that in Ref.~\cite{Mrowczynski:2016xqm}. 

With instantaneous coalescence in the rest frame of $pn$-pair or $ppn$-cluster, i.e., $\Delta t'=0$, we get the Lorentz transformation
\begin{eqnarray}
\boldsymbol{r} = \boldsymbol{r}' +(\gamma-1)\frac{\boldsymbol{r}'\cdot \boldsymbol{\beta}}{\beta^2}\boldsymbol{\beta},    \label{eq:LorentzTr}
\end{eqnarray}
where $\boldsymbol{\beta}$ is the three-dimensional velocity vector of the center-of-mass frame of $pn$-pair or $ppn$-cluster in the laboratory frame.
Substituting Eqs. (\ref{eq:fpn-rp}) and (\ref{eq:fppn-r1r2p}) into Eqs. (\ref{eq:fd-r}) and (\ref{eq:fHe3-r}) and using Eq. (\ref{eq:LorentzTr}) to integrate from relative coordinate variables, we can obtain 
{\setlength\arraycolsep{0.2pt}
\begin{eqnarray}
&& f_{d}(\boldsymbol{p}) = \frac{ 8 g_{d}(\sqrt{\pi}\hbar c)^3 \gamma}{\left[C R_f^2(\boldsymbol{p})+\sigma_d^2\right] \sqrt{C [R_f(\boldsymbol{p})/\gamma]^2+\sigma_d^2}} 
                       f_{pn}(\frac{\boldsymbol{p}}{2},\frac{\boldsymbol{p}}{2}) ,~~~~~   \label{eq:fd-pn}   \\
%--------------------------------------------------------------------------------------------------
&& f_{\mathrm{^{3}He}}(\boldsymbol{p}) = \frac{8^2 g_{\mathrm{^{3}He}} (\pi\hbar^2 c^2)^3 \gamma^2 }{3\sqrt{3}\left[\frac{C}{2} R_f^2(\boldsymbol{p})+\sigma_{\mathrm{^{3}He}}^2\right] \sqrt{\frac{C}{2} [R_f(\boldsymbol{p})/\gamma]^2+\sigma_{\mathrm{^{3}He}}^2} } \nonumber  \\
&&  ~~~~~~~~~~~~~~~  \times  \frac{1}{\left[\frac{2C}{3} R_f^2(\boldsymbol{p})+\sigma_{\mathrm{^{3}He}}^2\right] \sqrt{\frac{2C}{3} [R_f(\boldsymbol{p})/\gamma]^2+\sigma_{\mathrm{^{3}He}}^2}}   \nonumber  \\
 && ~~~~~~~~~~~~~~~   \times  f_{ppn}(\frac{\boldsymbol{p}}{3},\frac{\boldsymbol{p}}{3},\frac{\boldsymbol{p}}{3}).        \label{eq:fHe3-ppn}  
\end{eqnarray} }%

%_________________________________________________________________________________________ fd fHe 
Ignoring correlations between protons and neutrons, we have the three-dimensional momentum distributions of light nuclei as
{\setlength\arraycolsep{0.2pt}
\begin{eqnarray}
&& f_{d}(\boldsymbol{p}) = \frac{ 8 g_{d}(\sqrt{\pi}\hbar c)^3 \gamma}{\left[C R_f^2(\boldsymbol{p})+\sigma_d^2\right] \sqrt{C [R_f(\boldsymbol{p})/\gamma]^2+\sigma_d^2}}    
        f_{p}(\frac{\boldsymbol{p}}{2}) f_{n}(\frac{\boldsymbol{p}}{2}) , ~~~~~~    \label{eq:fd-final}  \\
%--------------------------------------------------------------------------------------------------
&& f_{\mathrm{^{3}He}}(\boldsymbol{p}) = \frac{8^2 g_{\mathrm{^{3}He}} (\pi\hbar^2 c^2)^3 \gamma^2 }{3\sqrt{3}\left[\frac{C}{2} R_f^2(\boldsymbol{p})+\sigma_{\mathrm{^{3}He}}^2\right] \sqrt{\frac{C}{2} [R_f(\boldsymbol{p})/\gamma]^2+\sigma_{\mathrm{^{3}He}}^2} } \nonumber  \\
&& ~~~~~~~~~~~~~~~~ \times  \frac{1}{\left[\frac{2C}{3} R_f^2(\boldsymbol{p})+\sigma_{\mathrm{^{3}He}}^2\right] \sqrt{\frac{2C}{3} [R_f(\boldsymbol{p})/\gamma]^2+\sigma_{\mathrm{^{3}He}}^2}}    \nonumber    \\
  && ~~~~~~~~~~~~~~~~ \times     f_{p}(\frac{\boldsymbol{p}}{3}) f_{p}(\frac{\boldsymbol{p}}{3}) f_{n}(\frac{\boldsymbol{p}}{3}). ~~~~~~    \label{eq:fHe3-final}  
\end{eqnarray} }%

From Eqs. (\ref{eq:fd-final}) and (\ref{eq:fHe3-final}), we can get the Lorentz-invariant momentum distributions of light nuclei. 
We denote the invariant distribution $\dfrac{d^{2}N}{2\pi p_{T}dp_{T}dy}$ with $f^{(inv)}$ and at the midrapidity $y=0$ we have
{\setlength\arraycolsep{0.2pt}
\begin{eqnarray}
&&  f_{d}^{(inv)}(p_{T}) = \frac{ 32 g_{d}(\sqrt{\pi}\hbar c)^3 }{m_{d}\left[C R_f^2(p_T)+\sigma_d^2\right] \sqrt{C [R_f(p_T)/\gamma]^2+\sigma_d^2}}   \nonumber \\
&& ~~~~~~~~~~~~~~~~~~~ \times f_{p}^{(inv)}(\frac{p_{T}}{2}) f_{n}^{(inv)}(\frac{p_{T}}{2}) ,   \label{eq:pt-d}          \\
%---------------------------------------------------------------------------------------------------------------------------------------------------------------------------
&&  f_{\mathrm{^{3}He}}^{(inv)}(p_{T}) = \frac{192\sqrt{3}g_{\mathrm{^{3}He}} (\pi\hbar^2 c^2)^3  }{m_{\mathrm{^{3}He}}^{2}\left[\frac{C}{2} R_f^2(p_T)+\sigma_{\mathrm{^{3}He}}^2\right]   \left[\frac{2C}{3} R_f^2(p_T)+\sigma_{\mathrm{^{3}He}}^2\right] } \nonumber  \\
&&~~~~~ \times  \frac{1}{ \sqrt{\frac{C}{2} [R_f(p_T)/\gamma]^2+\sigma_{\mathrm{^{3}He}}^2}  \sqrt{\frac{2C}{3} [R_f(p_T)/\gamma]^2+\sigma_{\mathrm{^{3}He}}^2}}      \nonumber  \\
&&~~~~~ \times  f_{p}^{(inv)}(\frac{p_{T}}{3}) f_{p}^{(inv)}(\frac{p_{T}}{3}) f_{n}^{(inv)}(\frac{p_{T}}{3}).       \label{eq:pt-He3}
\end{eqnarray} }%
Here $m_{d}$ is the mass of the $d$ and $m_{\mathrm{^{3}He}}$ is that of the $^3$He.
For tritons, we similarly have
{\setlength\arraycolsep{0.2pt}
\begin{eqnarray}
&&  f_{t}^{(inv)}(p_{T}) = \frac{192\sqrt{3} g_{t} (\pi\hbar^2 c^2)^3 }{m_{t}^{2}\left[\frac{C}{2} R_f^2(p_T)+\sigma_{t}^2\right]   \left[\frac{2C}{3} R_f^2(p_T)+\sigma_{t}^2\right] } \nonumber  \\
&&~~~~~~~~ \times  \frac{1}{ \sqrt{\frac{C}{2} [R_f(p_T)/\gamma]^2+\sigma_{t}^2}  \sqrt{\frac{2C}{3} [R_f(p_T)/\gamma]^2+\sigma_{t}^2}}      \nonumber  \\
&&~~~~~~~~ \times  f_{p}^{(inv)}(\frac{p_{T}}{3}) f_{n}^{(inv)}(\frac{p_{T}}{3}) f_{n}^{(inv)}(\frac{p_{T}}{3}),      \label{eq:pt-t}
\end{eqnarray} }%
where $g_{t}=1/4$ and $\sigma_{t}=R_{t}=1.7591$ fm~\cite{Angeli:2013epw}. $m_{t}$ is the mass of the $t$.
Eqs. (\ref{eq:pt-d}-\ref{eq:pt-t}) show relationships of light nuclei with primordial nucleons in momentum space in the laboratory frame. 
They can be used to calculate coalescence factors, yield rapidity densities and $p_T$ spectra of light nuclei in high energy collisions, 
especially in heavy ion collisions at the LHC where the coupling effect of coordinate and momentum may be intenser due to stronger collective motions and larger temperature gradients.
We will show their applications in Pb-Pb collisions at $\sqrt{s_{NN}}=2.76$ TeV in the following sections.

%-------------------------------------------------------------------------------------------------------------------------------------------------------------------------------------------------------------------------------------------------------- Results
\section{Results of coalescence factors}  \label{BA}

The coalescence factor $B_A$ is defined as
\begin{eqnarray}
 B_A  = f_{d,\mathrm{^{3}He},t}^{(inv)}(p_{T}) /  \left[    \left( f_{p}^{(inv)}(\frac{p_{T}}{A}) \right)^Z  \left( f_{n}^{(inv)}(\frac{p_{T}}{A}) \right)^{A-Z}   \right],  \label{eq:BA}
\end{eqnarray}
where $A$ is the mass number and $Z$ is the charge of the light nuclei.
$B_A$ is a key link between the formed light nuclei and the primordial nucleons, and folds important kinetic and dynamical information of the coalescence process.
Intuitively unfolding $B_A$ and a quantitative explanation for its centrality and $p_T$-dependent behaviors in heavy ion collisions at the LHC are necessary.

Substituting Eqs. (\ref{eq:pt-d}-\ref{eq:pt-t}) into Eq. (\ref{eq:BA}), we respectively have for $d$, $^3$He and $t$
{\setlength\arraycolsep{0.2pt}
\begin{eqnarray}
&&  B_2(p_{T}) = \frac{ 32 g_{d}(\sqrt{\pi}\hbar c)^3 }{m_{d}\left[C R_f^2(p_T)+\sigma_d^2\right] \sqrt{C [R_f(p_T)/\gamma]^2+\sigma_d^2}},    \label{eq:B2-d}          \\
%---------------------------------------------------------------------------------------------------------------------------------------------------------------------------
&&  B_3(p_{T}) = \frac{192\sqrt{3} g_{\mathrm{^{3}He}} (\pi\hbar^2 c^2)^3 }{m_{\mathrm{^{3}He}}^{2}\left[\frac{C}{2} R_f^2(p_T)+\sigma_{\mathrm{^{3}He}}^2\right]   \left[\frac{2C}{3} R_f^2(p_T)+\sigma_{\mathrm{^{3}He}}^2\right] } \nonumber  \\
&&~~ \times  \frac{1}{ \sqrt{\frac{C}{2} [R_f(p_T)/\gamma]^2+\sigma_{\mathrm{^{3}He}}^2}  \sqrt{\frac{2C}{3} [R_f(p_T)/\gamma]^2+\sigma_{\mathrm{^{3}He}}^2}}, ~~~~       \label{eq:B3-He3}  \\
%---------------------------------------------------------------------------------------------------------------------------------------------------------------------------
&&  B_3(p_{T}) = \frac{192\sqrt{3} g_{t} (\pi\hbar^2 c^2)^3 }{m_{t}^{2} \left[\frac{C}{2} R_f^2(p_T)+\sigma_{t}^2\right]   \left[\frac{2C}{3} R_f^2(p_T)+\sigma_{t}^2\right] } \nonumber  \\
&&~~ \times  \frac{1}{ \sqrt{\frac{C}{2} [R_f(p_T)/\gamma]^2+\sigma_{t}^2}  \sqrt{\frac{2C}{3} [R_f(p_T)/\gamma]^2+\sigma_{t}^2}}.          \label{eq:B3-t}
\end{eqnarray} }%
The above equations clearly show that $B_2$ and $B_3$ depend on the masses $m_{d,\mathrm{^{3}He},t}$, the spin degeneracy factors $g_{d,\mathrm{^{3}He},t}$ and the sizes of light nuclei via $\sigma_{d,\mathrm{^{3}He},t}$.
The Lorentz contraction factor $\gamma$, resulting from setting nucleon coalescence criteria in the rest frame of the nucleon pair or three-nucleon cluster rather than in the laboratory frame, affects the $p_T$-dependent behaviors of $B_2$ and $B_3$.
This has been studied in Ref.~\cite{Wang:2020zaw}.
The other influencing factor for $p_T$-dependent behaviors of $B_2$ and $B_3$ is the $R_f(p_{T})$, which is also closely related with centrality-dependent behaviors of $B_2$ and $B_3$.

\begin{figure}[!htb]
\includegraphics[width=\hsize]{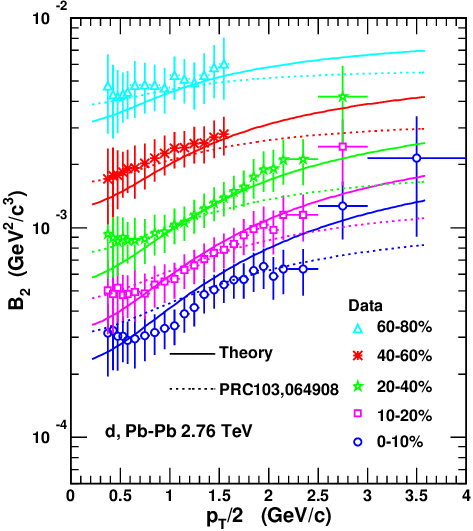}
\caption{The $B_2$ of $d$ as a function of $p_T/2$ in different centralities in Pb-Pb collisions at $\sqrt{s_{NN}}=2.76$ TeV.
Symbols with error bars are experimental data~\cite{ALICE:2015wav,ALICE:2017nuf} and different solid lines are theoretical results.
Different dotted lines are results with the coordinate-momentum factorization assumption in Ref.~\cite{Wang:2020zaw}.}
\label{fig:B2}
\end{figure}

\begin{figure*}[!htb]
\includegraphics[width=0.95\hsize]{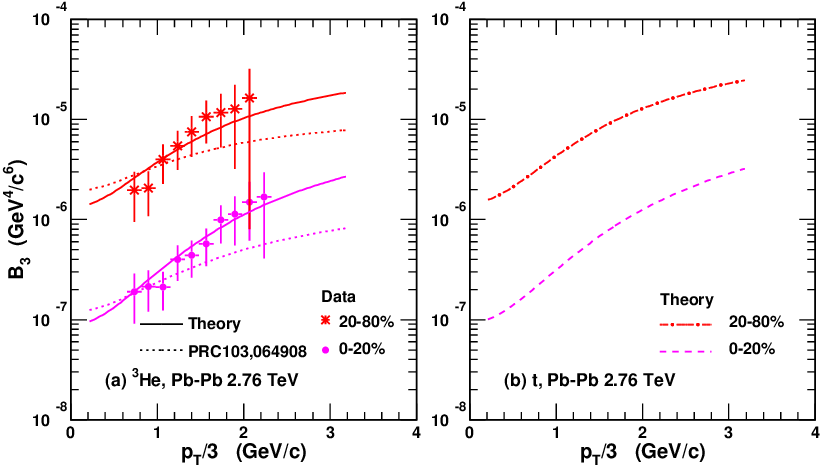}
\caption{The $B_3$ of (a) $^3$He and (b) $t$ as a function of $p_T/3$ in different centralities in Pb-Pb collisions at $\sqrt{s_{NN}}=2.76$ TeV.
Symbols with error bars are experimental data~\cite{ALICE:2015wav} and different solid, dashed and dash-dotted lines are our theoretical results.
Different dotted lines in panel (a) are results of $^3$He with the coordinate-momentum factorization assumption in Ref.~\cite{Wang:2020zaw}.}
\label{fig:B3}
\end{figure*}

To further compute $B_2$ and $B_3$, the specific form of $R_{f}(p_{T})$ is necessary.
In heavy ion collisions at CERN-SPS energies, it has been found that $R_{f}(p_{T})$ adopted as the femtoscopic radius can describe the $d$ production well~\cite{Scheibl:1998tk}.
If this still holds at LHC energies, the dependence of $R_{f}(p_{T})$ on centrality and $p_T$ should factorize into a linear dependence on the cube root of the pseudorapidity density of charged particles $(dN_{ch}/d\eta)^{1/3}$ and a power-law dependence on the transverse mass of the formed light nucleus $m_T$~\cite{ALICE:2015tra}.
So we get
{\setlength\arraycolsep{0.2pt}
\begin{eqnarray}
R_{f}(p_{T})=a*\left(\frac{dN_{ch}}{d\eta}\right)^{1/3}*\left(\sqrt{p_T^2+m_{d,\mathrm{^{3}He},t}^2}\right)^b, ~~~
\end{eqnarray} }%
where $a$ and $b$ are free parameters.
Their values in Pb-Pb collisions at $\sqrt{s_{NN}}=2.76$ TeV are (0.67,-0.25) for $d$ and (0.60,-0.25) for $^3$He and $t$, which are determined by reproducing the data of the $p_T$ spectra of $d$ in 0-10\% centrality and that of $^3$He in 0-20\% centrality.
Here $b$ is set to be centrality independent, which is consistent with that in hydrodynamics~\cite{Chakraborty:2020tym} and that in STAR measurements of two-pion interferometry in central and simi-central Au-Au collisions~\cite{STAR:2004qya}.
$a$ is also centrality independent.
Precise experimental measurements of HBT femtoscopic radius for nucleons in the future can crosscheck the scaling behaviors of $R_{f}$ as functions of $dN_{ch}/d\eta$ and $p_T$.

We use the data of $dN_{ch}/d\eta$ in Ref.~\cite{ALICE:2013mez} to get $R_f(p_T)$, and then compute $B_2$ and $B_3$.
Fig.~\ref{fig:B2} shows $B_2$ of $d$ as a function of the transverse momentum scaled by the mass number $p_T/2$ in different centralities in Pb-Pb collisions at $\sqrt{s_{NN}}=2.76$ TeV.
Symbols with error bars are experimental data~\cite{ALICE:2015wav,ALICE:2017nuf} and different solid lines are our theoretical results of the current nucleon coalescence model.
Different dotted lines are results from Ref.~\cite{Wang:2020zaw} where the assumption of the coordinate-momentum factorization was adopted.
From Fig.~\ref{fig:B2}, one can see from central to peripheral collisions, $B_2$ increases.
This is due to the decreasing scale of the hadronic system, which makes it easier for a $pn$-pair to recombine into a deuteron.
For the same centrality, $B_2$ increase as a function of $p_T/2$.
This increase behavior results on one hand from the Lorentz contraction factor $\gamma$~\cite{Wang:2020zaw}.
On the other hand, it results from the decreasing $R_f$ with increasing momentum.
The rising behavior of the experimental data as a function of $p_T/2$ from central to peripheral collisions can be quantitatively described by the current nucleon coalescence model. 

Fig.~\ref{fig:B3} (a) shows $B_3$ of $^3$He as a function of $p_T/3$ in different centralities in Pb-Pb collisions at $\sqrt{s_{NN}}=2.76$ TeV.
Symbols with error bars are experimental data~\cite{ALICE:2015wav} and different solid lines are our theoretical results.
Different dotted lines are results from Ref.~\cite{Wang:2020zaw} where the assumption of the coordinate-momentum factorization was adopted.
Similarly as $B_2$, experimental data of $B_3$ for $^3$He also exhibits a rising trend as a function of $p_T/3$, which is reproduced well by the current nucleon coalescence model from central to peripheral collisions.
Predictions of $B_3$ for $t$ in Fig.~\ref{fig:B3} (b) show similar trend as that of $^3$He, which can be tested by future experimental measurements.
Compared the current results denoted by solid lines with those in Ref.~\cite{Wang:2020zaw} denoted by dotted lines in Fig.~\ref{fig:B2} and Fig.~\ref{fig:B3} (a), one can see the improved nucleon coalescence model can better describe the slopes of $B_2$ and $B_3$.
%To quantitatively see the improvement, we introduce the averaged deviation degree $\delta_{devi}$. 
%It characterizes the global deviation extent of theoretical results from the data and is defined as follows: 
%{\setlength\arraycolsep{0pt}
%\begin{eqnarray}
% \delta_{devi} = \frac{1}{N_{\text{Data}}} \sum\limits_{j} \left| \frac{\text{Theory}_{j}-\text{Data}_{j}}{\text{Data}_{j}} \right| .   \label{eq:deviation}  
%\end{eqnarray} }%
%Take $^3$He in 0-20\% centrality as an example to execute the calculation. The value of $\delta_{devi}$ in the current nucleon coalescence model is 18\% and that in Ref.~\cite{Wang:2020zaw} where the assumption of the coordinate-momentum factorization was adopted is 34\%.
%This supports the necessary of including the coordinate-momentum correlation in heavy ion collisions at high LHC energies.

At the end of this section, we want to emphasize that the centrality and momentum dependent behaviors of $B_2$ and $B_3$ in Pb-Pb collisions at $\sqrt{s_{NN}}=2.76$ TeV are simultaneously explained by the improved nucleon coalescence model.
The influencing factors of $B_2$ and $B_3$ are explicitly unfolded, as shown in Eqs.~(\ref{eq:B2-d}-\ref{eq:B3-t}).
Some other models based on transport approach are also used to study behaviors of $B_A$ in heavy ion collisions at the high LHC energies~\cite{Oliinychenko:2018ugs,Oliinychenko:2018odl,Bailung:2023dpv,Liu:2022vbg}.
All the results from these different models can help cross understand production properties of light nuclei from different aspects.

%\textcolor{magenta}{We here use a unified approach based on the nucleon coalescence to explain the centrality and momentum dependent behaviors of $B_2$ and $B_3$ in Pb-Pb collisions at $\sqrt{s_{NN}}=2.76$ TeV.
%The influencing factors are explicitly unfolded in Eqs.~(\ref{eq:B2-d}-\ref{eq:B3-t}).
%Some other models based on popular transport approach are also employed to study behaviors of $B_A$ in heavy ion collisions at the high LHC energies.
%A hybrid method (hydrodynamics + transport) gave a good explanation for the centrality and momentum dependent behaviors of $B_2$ of $d$~\cite{Oliinychenko:2018ugs,Oliinychenko:2018odl}.
%A Multi-Phase Transport model with a coalescence afterburner gave quantitative reproductions for $B_2$ of $d$ and $B_3$ of $^3$He as a function of $p_T$ in central Pb-Pb collisions at $\sqrt{s_{NN}}=2.76$ TeV~\cite{Bailung:2023dpv}. 
%The PACIAE + DCPC model reproduced the increasing trend of $B_2$ of $d$ in peripheral $40–60$\% and central $0–5$\% Pb-Pb collisions at $\sqrt{s_{NN}}=2.76$ TeV~\cite{Liu:2022vbg}.
%All the results from these different models can help cross understanding production properties of light nuclei from different aspects.}

%--------------------------------------------------------------------------------------------------------------------------------------------------------------------------------------------------------------- pT spectra
\section{Results of $p_T$ spectra}  \label{pTspectra}

In this section, we use the nucleon coalescence model to study the $p_T$ spectra of light nuclei in different centralities in Pb-Pb collisions at $\sqrt{s_{NN}}=2.76$ TeV.
We first introduce the nucleon $p_T$ spectra. 
We then compute the $p_T$ spectra of $d$, $^3$He and $t$.
We finally calculate the averaged transverse momenta $\langle p_T \rangle$, the yield rapidity densities $dN/dy$ and  yield ratios of different light nuclei.

%------------------------------------------------------------------------------------------------------------------------------ nucleon pt spectra
\subsection{The $p_T$ spectra of primordial nucleons}

\begin{figure}[!htb]
\includegraphics[width=\hsize]{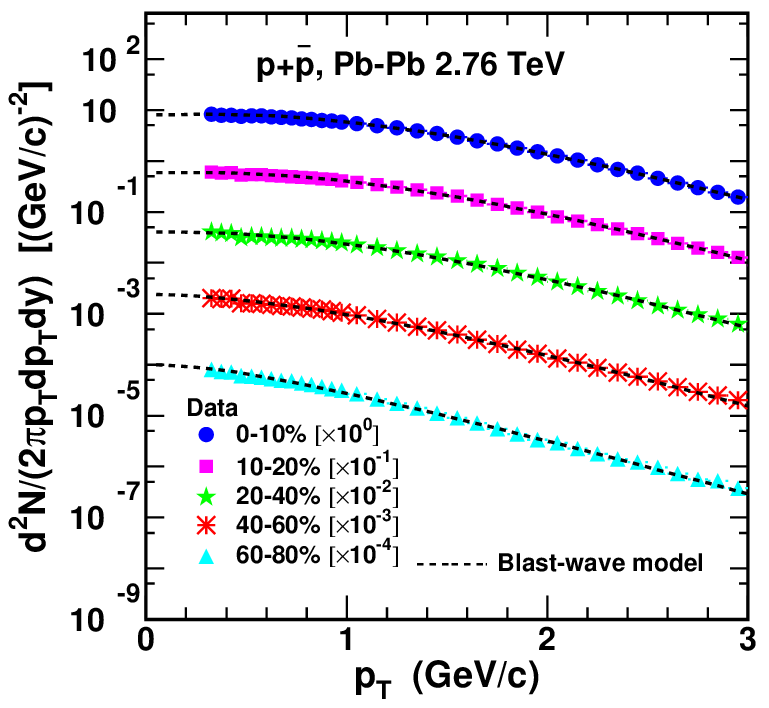}
\caption{The $p_T$ spectra of prompt protons plus antiprotons in different centralities in Pb-Pb collisions at $\sqrt{s_{NN}}=2.76$  TeV.
Symbols with error bars are experimental data~\cite{ALICE:2013mez}, and different lines are the results of the blast-wave model.}
\label{fig:ppt}
\end{figure}

The $p_T$ spectra of primordial nucleons are necessary inputs for computing $p_T$ distributions of light nuclei in the nucleon coalescence model.
We here use the blast-wave model to get $p_T$ distribution functions of primordial protons by fitting the experimental data of prompt (anti)protons in Ref.~\cite{ALICE:2013mez}.
The blast-wave function~\cite{Schnedermann:1993ws} is given as
{\setlength\arraycolsep{0.2pt}
\begin{eqnarray}
\frac{d^{2}N}{2\pi p_{T}dp_{T}dy}  \propto  &&  \int_{0}^{R} r dr m_T I_0 \left(\frac{p_Tsinh\rho}{T_{kin}}\right)   K_1\left(\frac{m_Tcosh\rho}{T_{kin}}\right), ~~~~~~  \label{eq:BWfitfunc}
\end{eqnarray} }%
where $r$ is the radial distance in the transverse plane and $R$ is the radius of the fireball.
$m_T$ is the transverse mass of the proton. $I_0$ and $K_1$ are the modified Bessel functions, and the velocity profile $\rho=tanh^{-1}[\beta_s(\frac{r}{R})^n]$.
The surface velocity $\beta_s$, the kinetic freeze-out temperature $T_{kin}$ and $n$ are fitting parameters.

Fig.~\ref{fig:ppt} shows the $p_T$ spectra of prompt protons plus antiprotons in different centralities in Pb-Pb collisions at $\sqrt{s_{NN}}= 2.76$ TeV.
Symbols with error bars are experimental data~\cite{ALICE:2013mez}, and different lines are the results of the blast-wave model.
The $p_T$ spectra in different centralities are scaled by different factors for clarity as shown in the figure.
For the primordial neutron $p_T$ spectra, we adopt the same as those of primordial protons as we focus on light nuclei production at midrapidity at so high LHC energy that the isospin symmetry is well satisfied.
We in the following use these nucleon results from the blast-wave model to compute the productions of different light nuclei.

%------------------------------------------------------------------------------------------------------------------------------ light nuclei pt spectra
\subsection{The $p_T$ spectra of light nuclei}

\begin{figure}[!htb]
\includegraphics[width=\hsize]{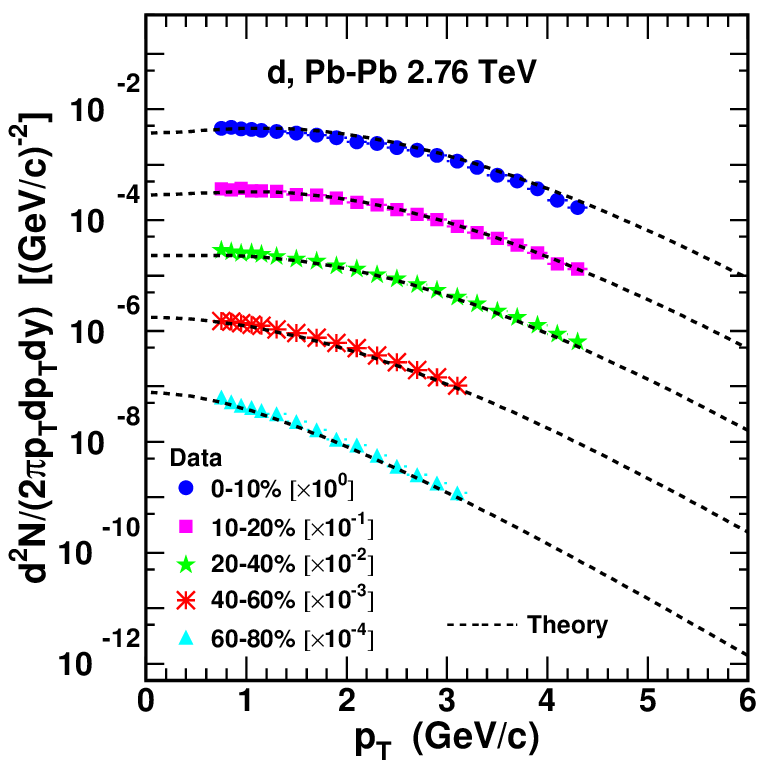}
\caption{The $p_T$ spectra of deuterons in different centralities in Pb-Pb collisions at $\sqrt{s_{NN}}=2.76$ TeV. 
Symbols are experimental data~\cite{ALICE:2015wav} and different lines are the theoretical results.}
\label{fig:dpt}
\end{figure}

With Eq.~(\ref{eq:pt-d}), we first calculate the $p_T$ spectra of deuterons in Pb-Pb collisions at $\sqrt{s_{NN}}=2.76$ TeV in $0-10$\%, $10-20$\%, $20-40$\%, $40-60$\% and $60-80$\% centralities.
Different lines scaled by different factors for clarity in Fig.~\ref{fig:dpt} are our theoretical results.
Symbols with error bars are experimental data from the ALICE collaboration~\cite{ALICE:2015wav}. 
From Fig.~\ref{fig:dpt}, one can see the $p+n$ coalescence can well reproduce the available data from central to peripheral Pb-Pb collisions at $\sqrt{s_{NN}}=2.76$ TeV.

\begin{figure*}[!htb]
\includegraphics[width=0.9\hsize]{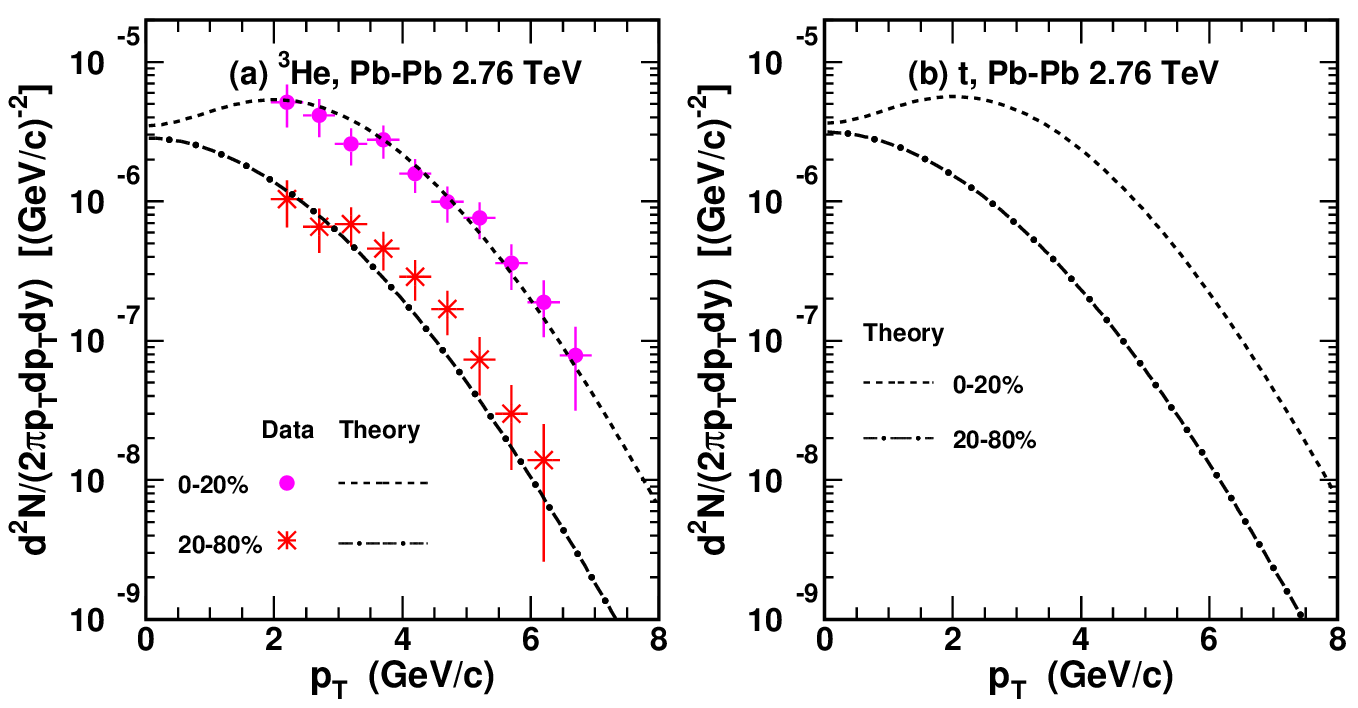}
\caption{The $p_T$ spectra of (a) $^3$He and (b) $t$ in different centralities in Pb-Pb collisions at $\sqrt{s_{NN}}=2.76$ TeV. 
Symbols are experimental data~\cite{ALICE:2015wav} and different lines are the theoretical results.}
\label{fig:He3tpt}
\end{figure*}

We then study the $p_T$ spectra of $^3$He and $t$ in Pb-Pb collisions at $\sqrt{s_{NN}}=2.76$ TeV in $0-20$\% and $20-80$\% centralities.
Different lines in Fig.~\ref{fig:He3tpt} (a) are our theoretical results of $^3$He, which agree with the available data denoted by filled symbols~\cite{ALICE:2015wav} within experimental uncertainties.
In low $p_T<2$ GeV/c region where the data are absent, our theoretical results show different trends in different centralities, slight increase in $0-20$\% centrality but decrease in $20-80$\% centrality.
This difference is caused by the competition of the $p_T$ distributions of nucleons and $R_f(p_T)$ in our model.
With the increase of the $p_T$, the decreasing nucleon $p_T$ distributions suppress $^3$He production while decreasing $R_f(p_T)$ enhances its production.
In central $0-20$\% collisions, nucleon $p_T$ distributions decrease very weakly or nearly hold invariant in $p_T<0.6$ GeV/c, so decreasing $R_f(p_T)$ as the function of $p_T$ makes the $p_T$ spectra of $^3$He increase in $p_T<2$ GeV/c.
In $20-80$\% centrality, although decreasing $R_f(p_T)$ still makes the $p_T$ spectra of $^3$He increase as the function of $p_T$, but obvious decreasing $p_T$ distributions of nucleons in $p_T<0.6$ GeV/c dominate the decreasing behavior of the $p_T$ spectra of $^3$He.
Future experimental measurements at low $p_T$ area can test the pattern of the $R_f(p_T)$ and the coalescence production mechanism for $^3$He.
Dashed line and dash-dotted line in Fig.~\ref{fig:He3tpt} (b) are predictions for $t$ in centralities $0-20$\% and $20-80$\%, respectively.

%------------------------------------------------------------------------------------------------------------------------------ light nuclei dN/dy & <pT>
\subsection{Averaged transverse momenta and yield rapidity densities of light nuclei}

\begin{table*}[!htb]
\begin{center}
\caption{Averaged transverse momenta $\langle p_T\rangle$ and yield rapidity densities $dN/dy$ of $d$, $^3$He and $t$ in different centralities in Pb-Pb collisions at $\sqrt{s_{NN}}=2.76$ TeV. Experimental data in the third and fifth columns are from Ref.~\cite{ALICE:2015wav}. Theoretical results are in the fourth and sixth columns.}  \label{table:dNdypt-dHe3t}
\begin{tabular}{@{\extracolsep{\fill}}cccccccc@{\extracolsep{\fill}}}
\toprule
     &\multirow{2}{*}{Centrality}     &\multicolumn{2}{@{}c@{}}{$\langle p_T\rangle$}          &          &\multicolumn{2}{@{}c@{}}{$dN/dy$} \\
\cline{3-4}  \cline{6-7} 
                          &     &Data                                 &Theory                   &          &Data                                                       &Theory      \\
\hline
\multirow{5}{*}{$d$}
  &0-10\%                  &$2.12\pm0.00\pm0.09$    &$2.19$                  &          &$(9.82\pm0.04\pm1.58)\times10^{-2}$  &$11.38\times10^{-2}$    \\
 &10-20\%                 &$2.07\pm0.01\pm0.10$    &$2.12$                  &          &$(7.60\pm0.04\pm1.25)\times10^{-2}$  &$7.55\times10^{-2}$    \\
 &20-40\%                 &$1.92\pm0.00\pm0.11$    &$1.95$                  &          &$(4.76\pm0.02\pm0.82)\times10^{-2}$  &$4.28\times10^{-2}$    \\
 &40-60\%                 &$1.63\pm0.01\pm0.09$    &$1.62$                  &          &$(1.90\pm0.01\pm0.41)\times10^{-2}$  &$1.71\times10^{-2}$    \\
 &60-80\%                 &$1.29\pm0.01\pm0.14$    &$1.28$                  &          &$(0.51\pm0.01\pm0.14)\times10^{-2}$  &$0.42\times10^{-2}$    \\
\hline
\multirow{2}{*}{$^3$He}
 &0-20\%                 &$2.83\pm0.05\pm0.45$    &$2.95$                  &          &$(2.76\pm0.09\pm0.62)\times10^{-4}$  &$2.60\times10^{-4}$    \\
 &20-80\%               &$2.65\pm0.06\pm0.45$    &$2.18$                  &          &$(5.09\pm0.24\pm1.36)\times10^{-5}$  &$5.14\times10^{-5}$    \\
\hline
\multirow{2}{*}{$t$}
 &0-20\%                 &$---$  &$2.97$      &          &$---$  &$2.77\times10^{-4}$    \\
 &20-80\%                 &$---$  &$2.20$      &          &$---$  &$5.84\times10^{-5}$    \\
\botrule
\end{tabular}
\end{center}
\end{table*}

We here study the averaged transverse momenta $\langle p_T\rangle$ and yield rapidity densities $dN/dy$ of $d$, $^3$He and $t$.
Our theoretical results are put in the fourth and sixth columns in Table \ref{table:dNdypt-dHe3t}.
Experimental data in the third and fifth columns are from Ref.~\cite{ALICE:2015wav}.
Theoretical results for $d$ and $^3$He are consistent with the corresponding data within the experimental uncertainties.
Predictions for $t$ are provided for future experimental measurements.
A clear decreasing trend for both $\langle p_T\rangle$ and $dN/dy$ from central to peripheral collisions is observed.
This is due to that in more central collisions more energy is deposited in the midrapidity region and collective evolution exists longer.

%------------------------------------------------------------------------------------------------------------------------------ Yield ratios of light nuclei
\subsection{Yield ratios of light nuclei}

\begin{figure*}[!htb]
\includegraphics[width=0.8\hsize]{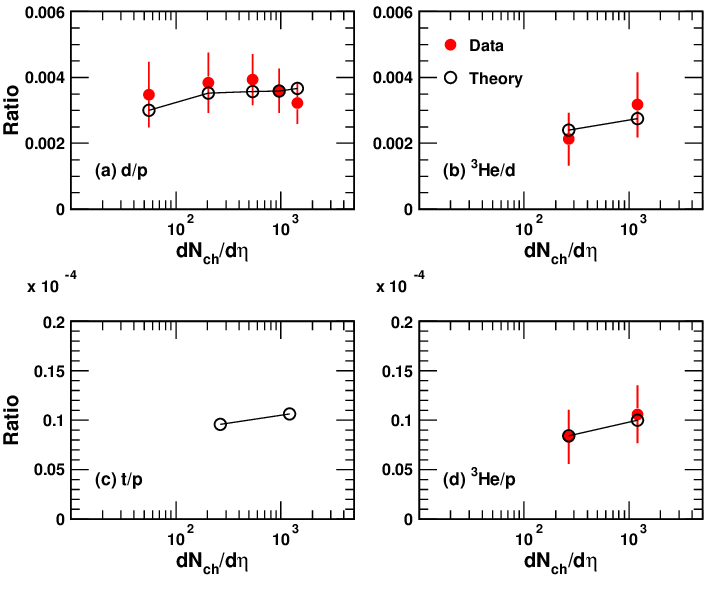}
\caption{Yield ratios (a) $d/p$, (b) $^3$He$/d$, (c) $t/p$ and (d) $^3$He$/p$ as a function of $dN_{ch}/d\eta$ in Pb-Pb collisions at $\sqrt{s_{NN}}=2.76$ TeV. 
Filled circles are experimental data~\cite{ALICE:2015wav} and open circles connected with solid lines to guide the eye are the theoretical results.}
\label{fig:Rdp}
\end{figure*}

\begin{figure*}[!htb]
\includegraphics[width=0.85\hsize]{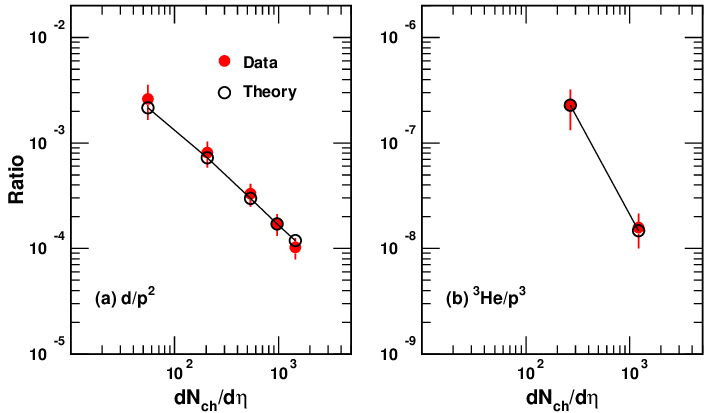}
\caption{Yield ratios (a) $d/p^2$ and (b) $^3$He$/p^3$ as a function of $dN_{ch}/d\eta$ in Pb-Pb collisions at $\sqrt{s_{NN}}=2.76$ TeV. 
Filled circles are experimental data~\cite{ALICE:2015wav} and open circles connected with solid lines to guide the eye are the theoretical results.}
\label{fig:Rdp2}
\end{figure*}

Yield ratios of light nuclei are characteristic probes for production mechanisms and contain intrinsic production correlations among different light nuclei.
In this subsection, we study three groups of yield ratios.
One is two-particle ratios such as $d/p$, $^3$He$/d$, $t/p$ and $^3$He$/p$.
The second group includes $d/p^{2}$ and $^3$He$/p^3$. They represent the probability of any nucleon-pair coalescing into a $d$ and that of  any $ppn$-cluster coalescing into a $^3$He.
The last is $t/^3$He, which exhibits interesting behaviors as functions of $p_T$ and the collision centrality.

From Eqs. (\ref{eq:pt-d}-\ref{eq:pt-t}) we approximately have the $p_T$-integrated yield ratios
{\setlength\arraycolsep{0.2pt}
\begin{eqnarray}
 \frac{d}{p}  \propto && \frac{ N_p }{\langle R_f \rangle^3 \left(C +\frac{\sigma_d^2}{\langle R_f \rangle^2}\right)  \sqrt{\frac{C}{\langle\gamma\rangle^2}+\frac{\sigma_d^2}{\langle R_f \rangle^2}}},  \label{eq:Rdp}         \\
%---------------------------------------------------------------------------------------------------------------------------------------------------------------------------
  \frac{^3\text{He}}{d} \propto 
&& \frac{ N_p  \left(C +\frac{\sigma_d^2}{\langle R_f \rangle^2} \right)  \sqrt{\frac{C}{\langle\gamma\rangle^2}+\frac{\sigma_d^2}{\langle R_f \rangle^2}}}
{\langle R_f \rangle^3 \left(\frac{C}{2} + \frac{\sigma_{\mathrm{^{3}He}}^2}{\langle R_f \rangle^2}\right)  \sqrt{\frac{C}{2\langle\gamma\rangle^2} +\frac{\sigma_{\mathrm{^{3}He}}^2}{\langle R_f \rangle^2}}   } \nonumber  \\
&& \times \frac{1}{  \left(\frac{2C}{3} + \frac{\sigma_{\mathrm{^{3}He}}^2}{\langle R_f \rangle^2} \right) \sqrt{\frac{2C}{3\langle\gamma\rangle^2}+\frac{\sigma_{\mathrm{^{3}He}}^2}{\langle R_f \rangle^2}}},   \nonumber  \\
\approx 
&&  \frac{2^{3/2}N_p}{ \langle R_f \rangle^3  \left(\frac{2C}{3} + \frac{\sigma_{\mathrm{^{3}He}}^2}{\langle R_f \rangle^2} \right) \sqrt{\frac{2C}{3\langle\gamma\rangle^2}+\frac{\sigma_{\mathrm{^{3}He}}^2}{\langle R_f \rangle^2}}}   \nonumber  \\
&&  \times\left\{1+\Delta\epsilon^2 \left[ \frac{1}{1+\frac{C\langle R_f \rangle^2}{(\sqrt{2}\sigma_{\mathrm{^{3}He}})^2}} +\frac{1/2}{1+\frac{C\langle R_f \rangle^2/\langle\gamma\rangle^2}{(\sqrt{2}\sigma_{\mathrm{^{3}He}})^2}}  \right]   \right\}, ~~       \label{eq:RHed}           \\
%---------------------------------------------------------------------------------------------------------------------------------------------------------------------------
 \frac{t}{p} \propto &&  \frac{ N_p^2  }
{\langle R_f \rangle^6 \left(\frac{C}{2} + \frac{\sigma_{t}^2}{\langle R_f \rangle^2}\right)   \left(\frac{2C}{3} + \frac{\sigma_{t}^2}{\langle R_f \rangle^2}\right)  }  \nonumber  \\
&& \times \frac{1}{\sqrt{\frac{C}{2\langle\gamma\rangle^2} +\frac{\sigma_{t}^2}{\langle R_f \rangle^2}}  \sqrt{\frac{2C}{3\langle\gamma\rangle^2}+\frac{\sigma_{t}^2}{\langle R_f \rangle^2}}},    \label{eq:Rtp}           \\
%---------------------------------------------------------------------------------------------------------------------------------------------------------------------------
 \frac{^3\text{He}}{p} \propto && \frac{ N_p^2  }
{\langle R_f \rangle^6 \left(\frac{C}{2} + \frac{\sigma_{\mathrm{^{3}He}}^2}{\langle R_f \rangle^2}\right)   \left(\frac{2C}{3} + \frac{\sigma_{\mathrm{^{3}He}}^2}{\langle R_f \rangle^2}\right)  }  \nonumber  \\
&& \times \frac{1}{\sqrt{\frac{C}{2\langle\gamma\rangle^2} +\frac{\sigma_{\mathrm{^{3}He}}^2}{\langle R_f \rangle^2}}  \sqrt{\frac{2C}{3\langle\gamma\rangle^2}+\frac{\sigma_{\mathrm{^{3}He}}^2}{\langle R_f \rangle^2}}}.    \label{eq:RHep}    
\end{eqnarray} }%
The angle brackets denote the averaged values.
Note that in the approximately equal sign in Eq.~(\ref{eq:RHed}), we ignore the difference of $\langle R_f\rangle$ and that of $\langle \gamma\rangle$ for $d$ and $^3$He and ignore the higher order terms of
$\Delta\epsilon^2$, where $\Delta\epsilon^2=[\sigma_d^2-(\sqrt{2}\sigma_{\mathrm{^{3}He}})^2]/(\sqrt{2}\sigma_{\mathrm{^{3}He}})^2 < 1$.
Eqs.~(\ref{eq:Rdp}-\ref{eq:RHep}) show that centrality-dependent behaviors of these two-particle ratios are closely related with the nucleon density ${ N_p }/{\langle R_f \rangle^3}$, ${\sigma_d}/{\langle R_f \rangle}$ and $\langle\gamma\rangle$.
From peripheral to central collisions, i.e., with the increasing $dN_{ch}/d\eta$, $\langle R_f \rangle$ and $\langle\gamma\rangle$ increase.
Suppressions on these ratios from $\sigma_{d,\mathrm{^{3}He},t}/\langle R_f \rangle$ and ${1}/{\langle\gamma\rangle^2}$ become weak, and this makes these ratios increase.
The nucleon density $N_p /\langle R_f \rangle^3$ decreases with increasing $dN_{ch}/d\eta$, which can be deduced from the trend of the yield ratio $p/\pi$~\cite{ALICE:2013mez}.
This leads to the decrease of these ratios.
The final behaviors of $d/p$, $^3$He$/d$, $t/p$ and $^3$He$/p$ as the function of the centrality dependence denoted by $dN_{ch}/d\eta$ depend on the competition effect from $\sigma_{d,\mathrm{^{3}He},t}/\langle R_f \rangle$, ${1}/{\langle\gamma\rangle^2}$ and $N_p /\langle R_f \rangle^3$.

Fig. \ref{fig:Rdp} shows the $dN_{ch}/d\eta$ dependence of $d/p$, $^3$He$/d$, $t/p$ and $^3$He$/p$ in Pb-Pb collisions at $\sqrt{s_{NN}}=2.76$ TeV. 
Filled circles are experimental data~\cite{ALICE:2015wav}.
Open circles connected with solid lines to guide the eye are our theoretical results.
Results of our model agree with the available data within the experimental uncertainties.
The conjunct effect from $N_p /\langle R_f \rangle^3$, ${\sigma_d}/{\langle R_f \rangle}$ and ${C}/{\langle\gamma\rangle^2}$ makes $d/p$ approximately unchanged as the function of the collision centrality, as shown in Fig. \ref{fig:Rdp} (a).
Fig. \ref{fig:Rdp} (b), (c) and (d) show $^3$He$/d$, $t/p$ and $^3$He$/p$ increase slightly as the function of $dN_{ch}/d\eta$. 
The large experimental uncertainties of the data~\cite{ALICE:2015wav} make it hard to give a final conclusion.
The canonical effect in the thermal model leading to a reduction in peripheral collisions and the baryon-antibaryon annihilations in the hybrid simulation leading to an additional suppression in central collisions for $d/p$ and $^3$He$/p$ have been discussed in Ref.~\cite{Reichert:2022mek}.
Discussions from different models give explanations of these ratios from different viewpoints.

From Eqs.~(\ref{eq:Rdp}) and (\ref{eq:RHep}), we immediately have
{\setlength\arraycolsep{0.2pt}
\begin{eqnarray}
 \frac{d}{p^2}  \propto && \frac{ 1 }{\langle R_f \rangle^3 \left(C +\frac{\sigma_d^2}{\langle R_f \rangle^2}\right)  \sqrt{\frac{C}{\langle\gamma\rangle^2}+\frac{\sigma_d^2}{\langle R_f \rangle^2}}},  \label{eq:Rdp2}         \\
%---------------------------------------------------------------------------------------------------------------------------------------------------------------------------
 \frac{^3\text{He}}{p^3} \propto && \frac{ 1  }
{\langle R_f \rangle^6 \left(\frac{C}{2} + \frac{\sigma_{\mathrm{^{3}He}}^2}{\langle R_f \rangle^2}\right)   \left(\frac{2C}{3} + \frac{\sigma_{\mathrm{^{3}He}}^2}{\langle R_f \rangle^2}\right)  }  \nonumber  \\
&& \times \frac{1}{\sqrt{\frac{C}{2\langle\gamma\rangle^2} +\frac{\sigma_{\mathrm{^{3}He}}^2}{\langle R_f \rangle^2}}  \sqrt{\frac{2C}{3\langle\gamma\rangle^2}+\frac{\sigma_{\mathrm{^{3}He}}^2}{\langle R_f \rangle^2}}}.    \label{eq:RHep3}    
\end{eqnarray} }%
They give an decreasing trend with the increasing $R_f$.
They do not depend on the absolute nucleon numbers or the nucleon rapidity densities.
Fig. \ref{fig:Rdp2} (a) and (b) show the ratios $d/p^{2}$ and $^3$He$/p^3$, respectively, in Pb-Pb collisions at $\sqrt{s_{NN}}=2.76$ TeV. 
Both  $d/p^{2}$ and $^3$He$/p^3$ show explicit decreasing trend with the increasing $dN_{ch}/d\eta$, which is very different from the previous $d/p$ and $^3$He$/p$. 
Recalling that $d/p^{2}$ and $^3$He$/p^3$ represent the probability of any nucleon-pair coalescing into a deuteron and that of  any $ppn$-cluster coalescing into a $^3$He.
This means that it is more difficult for any nucleon-pair or $ppn$-cluster to recombine into a deuteron or $^3$He in larger hadronic system produced in more cental collisions.

\begin{figure*}[!htb]
\includegraphics[width=0.85\hsize]{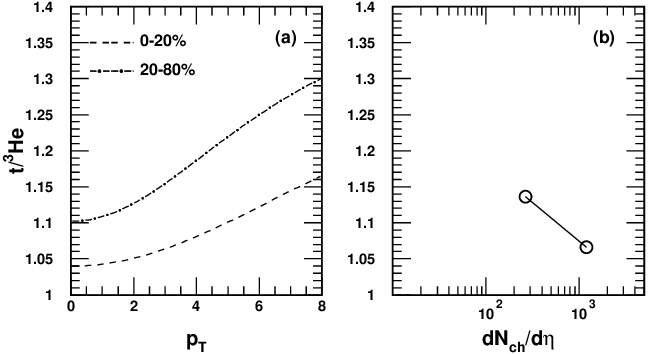}
\caption{Predictions of the yield ratio $t/^3$He as functions of (a) $p_T$ and (b) $dN_{ch}/d\eta$ in Pb-Pb collisions at $\sqrt{s_{NN}}=2.76$ TeV.}
\label{fig:RtHe3}
\end{figure*}

The yield ratio $t/^3$He provides another test of the coalescence productions for light nuclei.
With Eqs. (\ref{eq:pt-He3}) and (\ref{eq:pt-t}), we have its $p_T$-dependent function as
{\setlength\arraycolsep{0.2pt}
\begin{eqnarray}
%---------------------------------------------------------------------------------------------------------------------------------------------------------------------------
 \frac{t}{^3\text{He}}(p_T)  &=& \frac{   \left[\frac{C}{2} + \frac{\sigma_{\mathrm{^{3}He}}^2}{R_f ^2(p_T)}\right]   \left[\frac{2C}{3} + \frac{\sigma_{\mathrm{^{3}He}}^2}{R_f ^2(p_T)}\right]     }
{   \left[\frac{C}{2} + \frac{\sigma_{t}^2}{R_f ^2(p_T)}\right]   \left[\frac{2C}{3} + \frac{\sigma_{t}^2}{R_f ^2(p_T)}\right]    }  \nonumber  \\
&& \times  \frac{  \sqrt{\frac{C}{2\gamma^2} +\frac{\sigma_{\mathrm{^{3}He}}^2}{R_f ^2(p_T)}}  \sqrt{\frac{2C}{3\gamma^2}+\frac{\sigma_{\mathrm{^{3}He}}^2}{R_f ^2(p_T)}} }{   \sqrt{\frac{C}{2\gamma^2} +\frac{\sigma_{t}^2}{R_f ^2(p_T)}}  \sqrt{\frac{2C}{3\gamma^2}+\frac{\sigma_{t}^2}{R_f ^2(p_T)}} }  \nonumber  \\
& \approx & 1+ \frac{\Delta\sigma^2}{\sigma_t^2}   \left\{ \frac{1}{1+\frac{C}{2\sigma_t^2}R_f ^2(p_T)} + \frac{1}{1+\frac{2C}{3\sigma_t^2}R_f ^2(p_T)} \right.  \nonumber  \\
&&\left.    +\frac{1/2}{1+\frac{C}{2\sigma_t^2\gamma^2}R_f ^2(p_T)} + \frac{1/2}{1+\frac{2C}{3\sigma_t^2\gamma^2}R_f ^2(p_T)}  \right\}.      \label{eq:RtHe}      
\end{eqnarray} }%
Here $\Delta\sigma^2=\sigma^2_{\mathrm{^{3}He}}-\sigma^2_t$ and we ignore the higher order terms for the small quantity $\Delta\sigma^2/\sigma_t^2$.
Eq. (\ref{eq:RtHe}) shows that ${t}/{^3\text{He}}$ is always larger than one and approaches to one when $R_f \rightarrow \infty$.
The smaller $R_f $, the higher deviation of ${t}/{^3\text{He}}$ from one.
With the increasing $p_T$, $\gamma$ increases and $R_f$ decreases, so ${t}/{^3\text{He}}$ should increase.
Fig. \ref{fig:RtHe3} (a) shows our predictions of $t/^3$He as the function of $p_T$ in 0-20\% and 20-80\% centralities, respectively, in Pb-Pb collisions at $\sqrt{s_{NN}}=2.76$ TeV, both of which give increasing behaviors.
The $p_T$-integrated yield ratio ${t}/{^3\text{He}}$ as the function of $dN_{ch}/d\eta$ is in Fig. \ref{fig:RtHe3} (b), which has a decreasing trend.
This is because larger $dN_{ch}/d\eta$, i.e., larger $R_f$, makes ${t}/{^3\text{He}}$ decrease closer to one.
Predictions of $t/^3$He in the nucleon coalescence model give non-flat behaviors as functions of $p_T$ and $dN_{ch}/d\eta$.
This is due to different relative production suppression between $^3$He and $t$ at different hadronic system scales.
This feature is very different from that in the thermal model, where the expectation for this ratio is one~\cite{Andronic:2010qu}.
This can be used to distinguish production mechanisms of $^3$He and $t$.

%------------------------------------------------------------------------------------------------------------------------------------------------------------------------------ Summary
\section{Summary}  \label{Summary}

To get intuitive understandings of production properties of light nuclei in heavy ion collisions at the LHC, we improved a nucleon coalescence model analytically to include the coordinate-momentum correlation in nucleon joint distributions.
We derived the momentum distributions of $d$, $^3$He and $t$.
We obtained relationships of light nuclei with primordial nucleons in momentum space in the laboratory frame.
We gave formulas of coalescence factors $B_2$, $B_3$ and yield ratios $d/p$, $^3$He$/d$, $t/p$, $^3$He$/p$, $d/p^{2}$, $^3$He$/p^3$, $t/^3$He.

We applied the improved nucleon coalescence model to Pb-Pb collisions at $\sqrt{s_{NN}}=2.76$ TeV to study productions of different light nuclei.
We first investigated $B_2$ and $B_3$ and gave quantitative explanations for their interesting behaviors as functions of the collision centrality and the $p_T/A$.
We then studied the centrality dependence of the $p_T$ spectra, yield rapidity densities and averaged transverse momenta of $d$, $^3$He and $t$ with the $p_T$ distributions of kinetic freeze-out protons obtained from the blast-wave model.
We finally studied yield ratios $d/p$, $^3$He$/d$, $t/p$, $^3$He$/p$, $d/p^{2}$, $^3$He$/p^3$, $t/^3$He and discussed their behaviors as functions of the collision centrality and the $p_T$.
We found the nucleon coalescence model including the coordinate-momentum correlation can reproduce the experimental data available well.
We furthermore found the system effective radius obtained in the coalescence production of light nuclei exhibited similar behaviors to HBT interferometry radius.
We especially argued that nontrivial behaviors of yield ratios were valuable probes of production mechanisms of light nuclei.

%---------------------------------------------------------------------------------------------------------------------------------------------------------------------------- 
\section*{Acknowledgements}

This work was supported in part by the National Natural Science Foundation of China under Grants No. 12175115 and No. 12375074, the Natural Science Foundation of Shandong Province, China, under Grants No. ZR2020MA097,
and Higher Educational Youth Innovation Science and Technology Program of Shandong Province under Grants No. 2020KJJ004.

%---------------------------------------------------------------------------------------------------------------------------------------------------------------------------- 
\bibliographystyle{apsrev4-1}
\bibliography{myref}

\begin{thebibliography}{73}%
\makeatletter
\providecommand \@ifxundefined [1]{%
 \@ifx{#1\undefined}
}%
\providecommand \@ifnum [1]{%
 \ifnum #1\expandafter \@firstoftwo
 \else \expandafter \@secondoftwo
 \fi
}%
\providecommand \@ifx [1]{%
 \ifx #1\expandafter \@firstoftwo
 \else \expandafter \@secondoftwo
 \fi
}%
\providecommand \natexlab [1]{#1}%
\providecommand \enquote  [1]{``#1''}%
\providecommand \bibnamefont  [1]{#1}%
\providecommand \bibfnamefont [1]{#1}%
\providecommand \citenamefont [1]{#1}%
\providecommand \href@noop [0]{\@secondoftwo}%
\providecommand \href [0]{\begingroup \@sanitize@url \@href}%
\providecommand \@href[1]{\@@startlink{#1}\@@href}%
\providecommand \@@href[1]{\endgroup#1\@@endlink}%
\providecommand \@sanitize@url [0]{\catcode `\\12\catcode `\$12\catcode
  `\&12\catcode `\#12\catcode `\^12\catcode `\_12\catcode `\%12\relax}%
\providecommand \@@startlink[1]{}%
\providecommand \@@endlink[0]{}%
\providecommand \url  [0]{\begingroup\@sanitize@url \@url }%
\providecommand \@url [1]{\endgroup\@href {#1}{\urlprefix }}%
\providecommand \urlprefix  [0]{URL }%
\providecommand \Eprint [0]{\href }%
\providecommand \doibase [0]{http://dx.doi.org/}%
\providecommand \selectlanguage [0]{\@gobble}%
\providecommand \bibinfo  [0]{\@secondoftwo}%
\providecommand \bibfield  [0]{\@secondoftwo}%
\providecommand \translation [1]{[#1]}%
\providecommand \BibitemOpen [0]{}%
\providecommand \bibitemStop [0]{}%
\providecommand \bibitemNoStop [0]{.\EOS\space}%
\providecommand \EOS [0]{\spacefactor3000\relax}%
\providecommand \BibitemShut  [1]{\csname bibitem#1\endcsname}%
\let\auto@bib@innerbib\@empty
%</preamble>
\bibitem [{\citenamefont {Nagle}\ \emph {et~al.}(1994)\citenamefont {Nagle},
  \citenamefont {Kumar}, \citenamefont {Bennett}, \citenamefont {Diebold},
  \citenamefont {Pope}, \citenamefont {Sorge},\ and\ \citenamefont
  {Sullivan}}]{Nagle:1994hm}%
  \BibitemOpen
  \bibfield  {author} {\bibinfo {author} {\bibfnamefont {J.~L.}\ \bibnamefont
  {Nagle}}, \bibinfo {author} {\bibfnamefont {B.~S.}\ \bibnamefont {Kumar}},
  \bibinfo {author} {\bibfnamefont {M.~J.}\ \bibnamefont {Bennett}}, \bibinfo
  {author} {\bibfnamefont {G.~E.}\ \bibnamefont {Diebold}}, \bibinfo {author}
  {\bibfnamefont {J.~K.}\ \bibnamefont {Pope}}, \bibinfo {author}
  {\bibfnamefont {H.}~\bibnamefont {Sorge}}, \ and\ \bibinfo {author}
  {\bibfnamefont {J.~P.}\ \bibnamefont {Sullivan}},\ }\href {\doibase
  10.1103/PhysRevLett.73.1219} {\bibfield  {journal} {\bibinfo  {journal}
  {Phys. Rev. Lett.}\ }\textbf {\bibinfo {volume} {73}},\ \bibinfo {pages}
  {1219} (\bibinfo {year} {1994})}\BibitemShut {NoStop}%
\bibitem [{\citenamefont {Chen}\ \emph {et~al.}(2018)\citenamefont {Chen},
  \citenamefont {Keane}, \citenamefont {Ma}, \citenamefont {Tang},\ and\
  \citenamefont {Xu}}]{Chen:2018tnh}%
  \BibitemOpen
  \bibfield  {author} {\bibinfo {author} {\bibfnamefont {J.}~\bibnamefont
  {Chen}}, \bibinfo {author} {\bibfnamefont {D.}~\bibnamefont {Keane}},
  \bibinfo {author} {\bibfnamefont {Y.-G.}\ \bibnamefont {Ma}}, \bibinfo
  {author} {\bibfnamefont {A.}~\bibnamefont {Tang}}, \ and\ \bibinfo {author}
  {\bibfnamefont {Z.}~\bibnamefont {Xu}},\ }\href {\doibase
  10.1016/j.physrep.2018.07.002} {\bibfield  {journal} {\bibinfo  {journal}
  {Phys. Rept.}\ }\textbf {\bibinfo {volume} {760}},\ \bibinfo {pages} {1}
  (\bibinfo {year} {2018})},\ \Eprint {http://arxiv.org/abs/1808.09619}
  {arXiv:1808.09619 [nucl-ex]} \BibitemShut {NoStop}%
\bibitem [{\citenamefont {Blum}\ and\ \citenamefont
  {Takimoto}(2019)}]{Blum:2019suo}%
  \BibitemOpen
  \bibfield  {author} {\bibinfo {author} {\bibfnamefont {K.}~\bibnamefont
  {Blum}}\ and\ \bibinfo {author} {\bibfnamefont {M.}~\bibnamefont
  {Takimoto}},\ }\href {\doibase 10.1103/PhysRevC.99.044913} {\bibfield
  {journal} {\bibinfo  {journal} {Phys. Rev. C}\ }\textbf {\bibinfo {volume}
  {99}},\ \bibinfo {pages} {044913} (\bibinfo {year} {2019})},\ \Eprint
  {http://arxiv.org/abs/1901.07088} {arXiv:1901.07088 [nucl-th]} \BibitemShut
  {NoStop}%
\bibitem [{\citenamefont {Bazak}\ and\ \citenamefont
  {Mrowczynski}(2020)}]{Bazak:2020wjn}%
  \BibitemOpen
  \bibfield  {author} {\bibinfo {author} {\bibfnamefont {S.}~\bibnamefont
  {Bazak}}\ and\ \bibinfo {author} {\bibfnamefont {S.}~\bibnamefont
  {Mrowczynski}},\ }\href {\doibase 10.1140/epja/s10050-020-00198-6} {\bibfield
   {journal} {\bibinfo  {journal} {Eur. Phys. J. A}\ }\textbf {\bibinfo
  {volume} {56}},\ \bibinfo {pages} {193} (\bibinfo {year} {2020})},\ \Eprint
  {http://arxiv.org/abs/2001.11351} {arXiv:2001.11351 [nucl-th]} \BibitemShut
  {NoStop}%
\bibitem [{\citenamefont {Gutbrod}\ \emph {et~al.}(1976)\citenamefont
  {Gutbrod}, \citenamefont {Sandoval}, \citenamefont {Johansen}, \citenamefont
  {Poskanzer}, \citenamefont {Gosset}, \citenamefont {Meyer}, \citenamefont
  {Westfall},\ and\ \citenamefont {Stock}}]{Gutbrod:1976zzr}%
  \BibitemOpen
  \bibfield  {author} {\bibinfo {author} {\bibfnamefont {H.~H.}\ \bibnamefont
  {Gutbrod}}, \bibinfo {author} {\bibfnamefont {A.}~\bibnamefont {Sandoval}},
  \bibinfo {author} {\bibfnamefont {P.~J.}\ \bibnamefont {Johansen}}, \bibinfo
  {author} {\bibfnamefont {A.~M.}\ \bibnamefont {Poskanzer}}, \bibinfo {author}
  {\bibfnamefont {J.}~\bibnamefont {Gosset}}, \bibinfo {author} {\bibfnamefont
  {W.~G.}\ \bibnamefont {Meyer}}, \bibinfo {author} {\bibfnamefont {G.~D.}\
  \bibnamefont {Westfall}}, \ and\ \bibinfo {author} {\bibfnamefont
  {R.}~\bibnamefont {Stock}},\ }\href {\doibase 10.1103/PhysRevLett.37.667}
  {\bibfield  {journal} {\bibinfo  {journal} {Phys. Rev. Lett.}\ }\textbf
  {\bibinfo {volume} {37}},\ \bibinfo {pages} {667} (\bibinfo {year}
  {1976})}\BibitemShut {NoStop}%
\bibitem [{\citenamefont {Aichelin}(1991)}]{Aichelin:1991xy}%
  \BibitemOpen
  \bibfield  {author} {\bibinfo {author} {\bibfnamefont {J.}~\bibnamefont
  {Aichelin}},\ }\href {\doibase 10.1016/0370-1573(91)90094-3} {\bibfield
  {journal} {\bibinfo  {journal} {Phys. Rept.}\ }\textbf {\bibinfo {volume}
  {202}},\ \bibinfo {pages} {233} (\bibinfo {year} {1991})}\BibitemShut
  {NoStop}%
\bibitem [{\citenamefont {Andronic}\ \emph {et~al.}(2018)\citenamefont
  {Andronic}, \citenamefont {Braun-Munzinger}, \citenamefont {Redlich},\ and\
  \citenamefont {Stachel}}]{Andronic:2017pug}%
  \BibitemOpen
  \bibfield  {author} {\bibinfo {author} {\bibfnamefont {A.}~\bibnamefont
  {Andronic}}, \bibinfo {author} {\bibfnamefont {P.}~\bibnamefont
  {Braun-Munzinger}}, \bibinfo {author} {\bibfnamefont {K.}~\bibnamefont
  {Redlich}}, \ and\ \bibinfo {author} {\bibfnamefont {J.}~\bibnamefont
  {Stachel}},\ }\href {\doibase 10.1038/s41586-018-0491-6} {\bibfield
  {journal} {\bibinfo  {journal} {Nature}\ }\textbf {\bibinfo {volume} {561}},\
  \bibinfo {pages} {321} (\bibinfo {year} {2018})},\ \Eprint
  {http://arxiv.org/abs/1710.09425} {arXiv:1710.09425 [nucl-th]} \BibitemShut
  {NoStop}%
\bibitem [{\citenamefont {Bzdak}\ \emph {et~al.}(2020)\citenamefont {Bzdak},
  \citenamefont {Esumi}, \citenamefont {Koch}, \citenamefont {Liao},
  \citenamefont {Stephanov},\ and\ \citenamefont {Xu}}]{Bzdak:2019pkr}%
  \BibitemOpen
  \bibfield  {author} {\bibinfo {author} {\bibfnamefont {A.}~\bibnamefont
  {Bzdak}}, \bibinfo {author} {\bibfnamefont {S.}~\bibnamefont {Esumi}},
  \bibinfo {author} {\bibfnamefont {V.}~\bibnamefont {Koch}}, \bibinfo {author}
  {\bibfnamefont {J.}~\bibnamefont {Liao}}, \bibinfo {author} {\bibfnamefont
  {M.}~\bibnamefont {Stephanov}}, \ and\ \bibinfo {author} {\bibfnamefont
  {N.}~\bibnamefont {Xu}},\ }\href {\doibase 10.1016/j.physrep.2020.01.005}
  {\bibfield  {journal} {\bibinfo  {journal} {Phys. Rept.}\ }\textbf {\bibinfo
  {volume} {853}},\ \bibinfo {pages} {1} (\bibinfo {year} {2020})},\ \Eprint
  {http://arxiv.org/abs/1906.00936} {arXiv:1906.00936 [nucl-th]} \BibitemShut
  {NoStop}%
\bibitem [{\citenamefont {Sun}\ \emph {et~al.}(2017)\citenamefont {Sun},
  \citenamefont {Chen}, \citenamefont {Ko},\ and\ \citenamefont
  {Xu}}]{Sun:2017xrx}%
  \BibitemOpen
  \bibfield  {author} {\bibinfo {author} {\bibfnamefont {K.-J.}\ \bibnamefont
  {Sun}}, \bibinfo {author} {\bibfnamefont {L.-W.}\ \bibnamefont {Chen}},
  \bibinfo {author} {\bibfnamefont {C.~M.}\ \bibnamefont {Ko}}, \ and\ \bibinfo
  {author} {\bibfnamefont {Z.}~\bibnamefont {Xu}},\ }\href {\doibase
  10.1016/j.physletb.2017.09.056} {\bibfield  {journal} {\bibinfo  {journal}
  {Phys. Lett. B}\ }\textbf {\bibinfo {volume} {774}},\ \bibinfo {pages} {103}
  (\bibinfo {year} {2017})},\ \Eprint {http://arxiv.org/abs/1702.07620}
  {arXiv:1702.07620 [nucl-th]} \BibitemShut {NoStop}%
\bibitem [{\citenamefont {Sun}\ \emph {et~al.}(2018)\citenamefont {Sun},
  \citenamefont {Chen}, \citenamefont {Ko}, \citenamefont {Pu},\ and\
  \citenamefont {Xu}}]{Sun:2018jhg}%
  \BibitemOpen
  \bibfield  {author} {\bibinfo {author} {\bibfnamefont {K.-J.}\ \bibnamefont
  {Sun}}, \bibinfo {author} {\bibfnamefont {L.-W.}\ \bibnamefont {Chen}},
  \bibinfo {author} {\bibfnamefont {C.~M.}\ \bibnamefont {Ko}}, \bibinfo
  {author} {\bibfnamefont {J.}~\bibnamefont {Pu}}, \ and\ \bibinfo {author}
  {\bibfnamefont {Z.}~\bibnamefont {Xu}},\ }\href {\doibase
  10.1016/j.physletb.2018.04.035} {\bibfield  {journal} {\bibinfo  {journal}
  {Phys. Lett. B}\ }\textbf {\bibinfo {volume} {781}},\ \bibinfo {pages} {499}
  (\bibinfo {year} {2018})},\ \Eprint {http://arxiv.org/abs/1801.09382}
  {arXiv:1801.09382 [nucl-th]} \BibitemShut {NoStop}%
\bibitem [{\citenamefont {Luo}\ \emph {et~al.}(2020)\citenamefont {Luo},
  \citenamefont {Shi}, \citenamefont {Xu},\ and\ \citenamefont
  {Zhang}}]{Luo:2020pef}%
  \BibitemOpen
  \bibfield  {author} {\bibinfo {author} {\bibfnamefont {X.}~\bibnamefont
  {Luo}}, \bibinfo {author} {\bibfnamefont {S.}~\bibnamefont {Shi}}, \bibinfo
  {author} {\bibfnamefont {N.}~\bibnamefont {Xu}}, \ and\ \bibinfo {author}
  {\bibfnamefont {Y.}~\bibnamefont {Zhang}},\ }\href {\doibase
  10.3390/particles3020022} {\bibfield  {journal} {\bibinfo  {journal}
  {Particles}\ }\textbf {\bibinfo {volume} {3}},\ \bibinfo {pages} {278}
  (\bibinfo {year} {2020})},\ \Eprint {http://arxiv.org/abs/2004.00789}
  {arXiv:2004.00789 [nucl-ex]} \BibitemShut {NoStop}%
\bibitem [{\citenamefont {Junnarkar}\ and\ \citenamefont
  {Mathur}(2019)}]{Junnarkar:2019equ}%
  \BibitemOpen
  \bibfield  {author} {\bibinfo {author} {\bibfnamefont {P.}~\bibnamefont
  {Junnarkar}}\ and\ \bibinfo {author} {\bibfnamefont {N.}~\bibnamefont
  {Mathur}},\ }\href {\doibase 10.1103/PhysRevLett.123.162003} {\bibfield
  {journal} {\bibinfo  {journal} {Phys. Rev. Lett.}\ }\textbf {\bibinfo
  {volume} {123}},\ \bibinfo {pages} {162003} (\bibinfo {year} {2019})},\
  \Eprint {http://arxiv.org/abs/1906.06054} {arXiv:1906.06054 [hep-lat]}
  \BibitemShut {NoStop}%
\bibitem [{\citenamefont {Morita}\ \emph {et~al.}(2020)\citenamefont {Morita},
  \citenamefont {Gongyo}, \citenamefont {Hatsuda}, \citenamefont {Hyodo},
  \citenamefont {Kamiya},\ and\ \citenamefont {Ohnishi}}]{Morita:2019rph}%
  \BibitemOpen
  \bibfield  {author} {\bibinfo {author} {\bibfnamefont {K.}~\bibnamefont
  {Morita}}, \bibinfo {author} {\bibfnamefont {S.}~\bibnamefont {Gongyo}},
  \bibinfo {author} {\bibfnamefont {T.}~\bibnamefont {Hatsuda}}, \bibinfo
  {author} {\bibfnamefont {T.}~\bibnamefont {Hyodo}}, \bibinfo {author}
  {\bibfnamefont {Y.}~\bibnamefont {Kamiya}}, \ and\ \bibinfo {author}
  {\bibfnamefont {A.}~\bibnamefont {Ohnishi}},\ }\href {\doibase
  10.1103/PhysRevC.101.015201} {\bibfield  {journal} {\bibinfo  {journal}
  {Phys. Rev. C}\ }\textbf {\bibinfo {volume} {101}},\ \bibinfo {pages}
  {015201} (\bibinfo {year} {2020})},\ \Eprint
  {http://arxiv.org/abs/1908.05414} {arXiv:1908.05414 [nucl-th]} \BibitemShut
  {NoStop}%
\bibitem [{\citenamefont {Adler}\ \emph {et~al.}(2001)\citenamefont {Adler}
  \emph {et~al.}}]{STAR:2001pbk}%
  \BibitemOpen
  \bibfield  {author} {\bibinfo {author} {\bibfnamefont {C.}~\bibnamefont
  {Adler}} \emph {et~al.} (\bibinfo {collaboration} {STAR}),\ }\href {\doibase
  10.1103/PhysRevLett.87.262301} {\bibfield  {journal} {\bibinfo  {journal}
  {Phys. Rev. Lett.}\ }\textbf {\bibinfo {volume} {87}},\ \bibinfo {pages}
  {262301} (\bibinfo {year} {2001})},\ \bibinfo {note} {[Erratum:
  Phys.Rev.Lett. 87, 279902 (2001)]},\ \Eprint
  {http://arxiv.org/abs/nucl-ex/0108022} {arXiv:nucl-ex/0108022} \BibitemShut
  {NoStop}%
\bibitem [{\citenamefont {Afanasiev}\ \emph {et~al.}(2007)\citenamefont
  {Afanasiev} \emph {et~al.}}]{PHENIX:2007tef}%
  \BibitemOpen
  \bibfield  {author} {\bibinfo {author} {\bibfnamefont {S.}~\bibnamefont
  {Afanasiev}} \emph {et~al.} (\bibinfo {collaboration} {PHENIX}),\ }\href
  {\doibase 10.1103/PhysRevLett.99.052301} {\bibfield  {journal} {\bibinfo
  {journal} {Phys. Rev. Lett.}\ }\textbf {\bibinfo {volume} {99}},\ \bibinfo
  {pages} {052301} (\bibinfo {year} {2007})},\ \Eprint
  {http://arxiv.org/abs/nucl-ex/0703024} {arXiv:nucl-ex/0703024} \BibitemShut
  {NoStop}%
\bibitem [{\citenamefont {Anticic}\ \emph {et~al.}(2016)\citenamefont {Anticic}
  \emph {et~al.}}]{NA49:2016qvu}%
  \BibitemOpen
  \bibfield  {author} {\bibinfo {author} {\bibfnamefont {T.}~\bibnamefont
  {Anticic}} \emph {et~al.} (\bibinfo {collaboration} {NA49}),\ }\href
  {\doibase 10.1103/PhysRevC.94.044906} {\bibfield  {journal} {\bibinfo
  {journal} {Phys. Rev. C}\ }\textbf {\bibinfo {volume} {94}},\ \bibinfo
  {pages} {044906} (\bibinfo {year} {2016})},\ \Eprint
  {http://arxiv.org/abs/1606.04234} {arXiv:1606.04234 [nucl-ex]} \BibitemShut
  {NoStop}%
\bibitem [{\citenamefont {Albergo}\ \emph {et~al.}(2002)\citenamefont {Albergo}
  \emph {et~al.}}]{Albergo:2002gi}%
  \BibitemOpen
  \bibfield  {author} {\bibinfo {author} {\bibfnamefont {S.}~\bibnamefont
  {Albergo}} \emph {et~al.},\ }\href {\doibase 10.1103/PhysRevC.65.034907}
  {\bibfield  {journal} {\bibinfo  {journal} {Phys. Rev. C}\ }\textbf {\bibinfo
  {volume} {65}},\ \bibinfo {pages} {034907} (\bibinfo {year}
  {2002})}\BibitemShut {NoStop}%
\bibitem [{\citenamefont {Adam}\ \emph
  {et~al.}(2016{\natexlab{a}})\citenamefont {Adam} \emph
  {et~al.}}]{ALICE:2015wav}%
  \BibitemOpen
  \bibfield  {author} {\bibinfo {author} {\bibfnamefont {J.}~\bibnamefont
  {Adam}} \emph {et~al.} (\bibinfo {collaboration} {ALICE}),\ }\href {\doibase
  10.1103/PhysRevC.93.024917} {\bibfield  {journal} {\bibinfo  {journal} {Phys.
  Rev. C}\ }\textbf {\bibinfo {volume} {93}},\ \bibinfo {pages} {024917}
  (\bibinfo {year} {2016}{\natexlab{a}})},\ \Eprint
  {http://arxiv.org/abs/1506.08951} {arXiv:1506.08951 [nucl-ex]} \BibitemShut
  {NoStop}%
\bibitem [{\citenamefont {Acharya}\ \emph {et~al.}(2020)\citenamefont {Acharya}
  \emph {et~al.}}]{ALICE:2020chv}%
  \BibitemOpen
  \bibfield  {author} {\bibinfo {author} {\bibfnamefont {S.}~\bibnamefont
  {Acharya}} \emph {et~al.} (\bibinfo {collaboration} {ALICE}),\ }\href
  {\doibase 10.1103/PhysRevC.102.055203} {\bibfield  {journal} {\bibinfo
  {journal} {Phys. Rev. C}\ }\textbf {\bibinfo {volume} {102}},\ \bibinfo
  {pages} {055203} (\bibinfo {year} {2020})},\ \Eprint
  {http://arxiv.org/abs/2005.14639} {arXiv:2005.14639 [nucl-ex]} \BibitemShut
  {NoStop}%
\bibitem [{\citenamefont {Adamczyk}\ \emph {et~al.}(2016)\citenamefont
  {Adamczyk} \emph {et~al.}}]{STAR:2016ydv}%
  \BibitemOpen
  \bibfield  {author} {\bibinfo {author} {\bibfnamefont {L.}~\bibnamefont
  {Adamczyk}} \emph {et~al.} (\bibinfo {collaboration} {STAR}),\ }\href
  {\doibase 10.1103/PhysRevC.94.034908} {\bibfield  {journal} {\bibinfo
  {journal} {Phys. Rev. C}\ }\textbf {\bibinfo {volume} {94}},\ \bibinfo
  {pages} {034908} (\bibinfo {year} {2016})},\ \Eprint
  {http://arxiv.org/abs/1601.07052} {arXiv:1601.07052 [nucl-ex]} \BibitemShut
  {NoStop}%
\bibitem [{\citenamefont {Adam}\ \emph {et~al.}(2020)\citenamefont {Adam} \emph
  {et~al.}}]{STAR:2020hya}%
  \BibitemOpen
  \bibfield  {author} {\bibinfo {author} {\bibfnamefont {J.}~\bibnamefont
  {Adam}} \emph {et~al.} (\bibinfo {collaboration} {STAR}),\ }\href {\doibase
  10.1103/PhysRevC.102.044906} {\bibfield  {journal} {\bibinfo  {journal}
  {Phys. Rev. C}\ }\textbf {\bibinfo {volume} {102}},\ \bibinfo {pages}
  {044906} (\bibinfo {year} {2020})},\ \Eprint
  {http://arxiv.org/abs/2007.04609} {arXiv:2007.04609 [nucl-ex]} \BibitemShut
  {NoStop}%
\bibitem [{\citenamefont {Zhang}(2021)}]{Zhang:2020ewj}%
  \BibitemOpen
  \bibfield  {author} {\bibinfo {author} {\bibfnamefont {D.}~\bibnamefont
  {Zhang}} (\bibinfo {collaboration} {STAR}),\ }\href {\doibase
  10.1016/j.nuclphysa.2020.121825} {\bibfield  {journal} {\bibinfo  {journal}
  {Nucl. Phys. A}\ }\textbf {\bibinfo {volume} {1005}},\ \bibinfo {pages}
  {121825} (\bibinfo {year} {2021})},\ \Eprint
  {http://arxiv.org/abs/2002.10677} {arXiv:2002.10677 [nucl-ex]} \BibitemShut
  {NoStop}%
\bibitem [{\citenamefont {Adam}\ \emph {et~al.}(2019)\citenamefont {Adam} \emph
  {et~al.}}]{STAR:2019sjh}%
  \BibitemOpen
  \bibfield  {author} {\bibinfo {author} {\bibfnamefont {J.}~\bibnamefont
  {Adam}} \emph {et~al.} (\bibinfo {collaboration} {STAR}),\ }\href {\doibase
  10.1103/PhysRevC.99.064905} {\bibfield  {journal} {\bibinfo  {journal} {Phys.
  Rev. C}\ }\textbf {\bibinfo {volume} {99}},\ \bibinfo {pages} {064905}
  (\bibinfo {year} {2019})},\ \Eprint {http://arxiv.org/abs/1903.11778}
  {arXiv:1903.11778 [nucl-ex]} \BibitemShut {NoStop}%
\bibitem [{\citenamefont {Abdulhamid}\ \emph {et~al.}(2023)\citenamefont
  {Abdulhamid} \emph {et~al.}}]{STAR:2022hbp}%
  \BibitemOpen
  \bibfield  {author} {\bibinfo {author} {\bibfnamefont {M.}~\bibnamefont
  {Abdulhamid}} \emph {et~al.} (\bibinfo {collaboration} {STAR}),\ }\href
  {\doibase 10.1103/PhysRevLett.130.202301} {\bibfield  {journal} {\bibinfo
  {journal} {Phys. Rev. Lett.}\ }\textbf {\bibinfo {volume} {130}},\ \bibinfo
  {pages} {202301} (\bibinfo {year} {2023})},\ \Eprint
  {http://arxiv.org/abs/2209.08058} {arXiv:2209.08058 [nucl-ex]} \BibitemShut
  {NoStop}%
\bibitem [{\citenamefont {Braun-Munzinger}\ and\ \citenamefont
  {D\"onigus}(2019)}]{Braun-Munzinger:2018hat}%
  \BibitemOpen
  \bibfield  {author} {\bibinfo {author} {\bibfnamefont {P.}~\bibnamefont
  {Braun-Munzinger}}\ and\ \bibinfo {author} {\bibfnamefont {B.}~\bibnamefont
  {D\"onigus}},\ }\href {\doibase 10.1016/j.nuclphysa.2019.02.006} {\bibfield
  {journal} {\bibinfo  {journal} {Nucl. Phys. A}\ }\textbf {\bibinfo {volume}
  {987}},\ \bibinfo {pages} {144} (\bibinfo {year} {2019})},\ \Eprint
  {http://arxiv.org/abs/1809.04681} {arXiv:1809.04681 [nucl-ex]} \BibitemShut
  {NoStop}%
\bibitem [{\citenamefont {Oliinychenko}(2021)}]{Oliinychenko:2020ply}%
  \BibitemOpen
  \bibfield  {author} {\bibinfo {author} {\bibfnamefont {D.}~\bibnamefont
  {Oliinychenko}},\ }\href {\doibase 10.1016/j.nuclphysa.2020.121754}
  {\bibfield  {journal} {\bibinfo  {journal} {Nucl. Phys. A}\ }\textbf
  {\bibinfo {volume} {1005}},\ \bibinfo {pages} {121754} (\bibinfo {year}
  {2021})},\ \Eprint {http://arxiv.org/abs/2003.05476} {arXiv:2003.05476
  [hep-ph]} \BibitemShut {NoStop}%
\bibitem [{\citenamefont {Dover}\ \emph {et~al.}(1991)\citenamefont {Dover},
  \citenamefont {Heinz}, \citenamefont {Schnedermann},\ and\ \citenamefont
  {Zimanyi}}]{Dover:1991zn}%
  \BibitemOpen
  \bibfield  {author} {\bibinfo {author} {\bibfnamefont {C.~B.}\ \bibnamefont
  {Dover}}, \bibinfo {author} {\bibfnamefont {U.~W.}\ \bibnamefont {Heinz}},
  \bibinfo {author} {\bibfnamefont {E.}~\bibnamefont {Schnedermann}}, \ and\
  \bibinfo {author} {\bibfnamefont {J.}~\bibnamefont {Zimanyi}},\ }\href
  {\doibase 10.1103/PhysRevC.44.1636} {\bibfield  {journal} {\bibinfo
  {journal} {Phys. Rev. C}\ }\textbf {\bibinfo {volume} {44}},\ \bibinfo
  {pages} {1636} (\bibinfo {year} {1991})}\BibitemShut {NoStop}%
\bibitem [{\citenamefont {Chen}\ \emph
  {et~al.}(2003{\natexlab{a}})\citenamefont {Chen}, \citenamefont {Ko},\ and\
  \citenamefont {Li}}]{Chen:2003qj}%
  \BibitemOpen
  \bibfield  {author} {\bibinfo {author} {\bibfnamefont {L.-W.}\ \bibnamefont
  {Chen}}, \bibinfo {author} {\bibfnamefont {C.~M.}\ \bibnamefont {Ko}}, \ and\
  \bibinfo {author} {\bibfnamefont {B.-A.}\ \bibnamefont {Li}},\ }\href
  {\doibase 10.1103/PhysRevC.68.017601} {\bibfield  {journal} {\bibinfo
  {journal} {Phys. Rev. C}\ }\textbf {\bibinfo {volume} {68}},\ \bibinfo
  {pages} {017601} (\bibinfo {year} {2003}{\natexlab{a}})},\ \Eprint
  {http://arxiv.org/abs/nucl-th/0302068} {arXiv:nucl-th/0302068} \BibitemShut
  {NoStop}%
\bibitem [{\citenamefont {Mrowczynski}(2020)}]{Mrowczynski:2020ugu}%
  \BibitemOpen
  \bibfield  {author} {\bibinfo {author} {\bibfnamefont {S.}~\bibnamefont
  {Mrowczynski}},\ }\href {\doibase 10.1140/epjst/e2020-000067-0} {\bibfield
  {journal} {\bibinfo  {journal} {Eur. Phys. J. ST}\ }\textbf {\bibinfo
  {volume} {229}},\ \bibinfo {pages} {3559} (\bibinfo {year} {2020})},\ \Eprint
  {http://arxiv.org/abs/2004.07029} {arXiv:2004.07029 [nucl-th]} \BibitemShut
  {NoStop}%
\bibitem [{\citenamefont {Andronic}\ \emph {et~al.}(2011)\citenamefont
  {Andronic}, \citenamefont {Braun-Munzinger}, \citenamefont {Stachel},\ and\
  \citenamefont {Stocker}}]{Andronic:2010qu}%
  \BibitemOpen
  \bibfield  {author} {\bibinfo {author} {\bibfnamefont {A.}~\bibnamefont
  {Andronic}}, \bibinfo {author} {\bibfnamefont {P.}~\bibnamefont
  {Braun-Munzinger}}, \bibinfo {author} {\bibfnamefont {J.}~\bibnamefont
  {Stachel}}, \ and\ \bibinfo {author} {\bibfnamefont {H.}~\bibnamefont
  {Stocker}},\ }\href {\doibase 10.1016/j.physletb.2011.01.053} {\bibfield
  {journal} {\bibinfo  {journal} {Phys. Lett. B}\ }\textbf {\bibinfo {volume}
  {697}},\ \bibinfo {pages} {203} (\bibinfo {year} {2011})},\ \Eprint
  {http://arxiv.org/abs/1010.2995} {arXiv:1010.2995 [nucl-th]} \BibitemShut
  {NoStop}%
\bibitem [{\citenamefont {D\"onigus}\ \emph {et~al.}(2022)\citenamefont
  {D\"onigus}, \citenamefont {R\"opke},\ and\ \citenamefont
  {Blaschke}}]{Donigus:2022haq}%
  \BibitemOpen
  \bibfield  {author} {\bibinfo {author} {\bibfnamefont {B.}~\bibnamefont
  {D\"onigus}}, \bibinfo {author} {\bibfnamefont {G.}~\bibnamefont {R\"opke}},
  \ and\ \bibinfo {author} {\bibfnamefont {D.}~\bibnamefont {Blaschke}},\
  }\href {\doibase 10.1103/PhysRevC.106.044908} {\bibfield  {journal} {\bibinfo
   {journal} {Phys. Rev. C}\ }\textbf {\bibinfo {volume} {106}},\ \bibinfo
  {pages} {044908} (\bibinfo {year} {2022})},\ \Eprint
  {http://arxiv.org/abs/2206.10376} {arXiv:2206.10376 [nucl-th]} \BibitemShut
  {NoStop}%
\bibitem [{\citenamefont {Mekjian}(1977)}]{Mekjian:1977ei}%
  \BibitemOpen
  \bibfield  {author} {\bibinfo {author} {\bibfnamefont {A.}~\bibnamefont
  {Mekjian}},\ }\href {\doibase 10.1103/PhysRevLett.38.640} {\bibfield
  {journal} {\bibinfo  {journal} {Phys. Rev. Lett.}\ }\textbf {\bibinfo
  {volume} {38}},\ \bibinfo {pages} {640} (\bibinfo {year} {1977})}\BibitemShut
  {NoStop}%
\bibitem [{\citenamefont {Siemens}\ and\ \citenamefont
  {Kapusta}(1979)}]{Siemens:1979dz}%
  \BibitemOpen
  \bibfield  {author} {\bibinfo {author} {\bibfnamefont {P.~J.}\ \bibnamefont
  {Siemens}}\ and\ \bibinfo {author} {\bibfnamefont {J.~I.}\ \bibnamefont
  {Kapusta}},\ }\href {\doibase 10.1103/PhysRevLett.43.1486} {\bibfield
  {journal} {\bibinfo  {journal} {Phys. Rev. Lett.}\ }\textbf {\bibinfo
  {volume} {43}},\ \bibinfo {pages} {1486} (\bibinfo {year}
  {1979})}\BibitemShut {NoStop}%
\bibitem [{\citenamefont {Cleymans}\ \emph {et~al.}(2011)\citenamefont
  {Cleymans}, \citenamefont {Kabana}, \citenamefont {Kraus}, \citenamefont
  {Oeschler}, \citenamefont {Redlich},\ and\ \citenamefont
  {Sharma}}]{Cleymans:2011pe}%
  \BibitemOpen
  \bibfield  {author} {\bibinfo {author} {\bibfnamefont {J.}~\bibnamefont
  {Cleymans}}, \bibinfo {author} {\bibfnamefont {S.}~\bibnamefont {Kabana}},
  \bibinfo {author} {\bibfnamefont {I.}~\bibnamefont {Kraus}}, \bibinfo
  {author} {\bibfnamefont {H.}~\bibnamefont {Oeschler}}, \bibinfo {author}
  {\bibfnamefont {K.}~\bibnamefont {Redlich}}, \ and\ \bibinfo {author}
  {\bibfnamefont {N.}~\bibnamefont {Sharma}},\ }\href {\doibase
  10.1103/PhysRevC.84.054916} {\bibfield  {journal} {\bibinfo  {journal} {Phys.
  Rev. C}\ }\textbf {\bibinfo {volume} {84}},\ \bibinfo {pages} {054916}
  (\bibinfo {year} {2011})},\ \Eprint {http://arxiv.org/abs/1105.3719}
  {arXiv:1105.3719 [hep-ph]} \BibitemShut {NoStop}%
\bibitem [{\citenamefont {Cai}\ \emph {et~al.}(2019)\citenamefont {Cai},
  \citenamefont {Cohen}, \citenamefont {Gelman},\ and\ \citenamefont
  {Yamauchi}}]{Cai:2019jtk}%
  \BibitemOpen
  \bibfield  {author} {\bibinfo {author} {\bibfnamefont {Y.}~\bibnamefont
  {Cai}}, \bibinfo {author} {\bibfnamefont {T.~D.}\ \bibnamefont {Cohen}},
  \bibinfo {author} {\bibfnamefont {B.~A.}\ \bibnamefont {Gelman}}, \ and\
  \bibinfo {author} {\bibfnamefont {Y.}~\bibnamefont {Yamauchi}},\ }\href
  {\doibase 10.1103/PhysRevC.100.024911} {\bibfield  {journal} {\bibinfo
  {journal} {Phys. Rev. C}\ }\textbf {\bibinfo {volume} {100}},\ \bibinfo
  {pages} {024911} (\bibinfo {year} {2019})},\ \Eprint
  {http://arxiv.org/abs/1905.02753} {arXiv:1905.02753 [nucl-th]} \BibitemShut
  {NoStop}%
\bibitem [{\citenamefont {Schwarzschild}\ and\ \citenamefont
  {Zupancic}(1963)}]{Schwarzschild:1963zz}%
  \BibitemOpen
  \bibfield  {author} {\bibinfo {author} {\bibfnamefont {A.}~\bibnamefont
  {Schwarzschild}}\ and\ \bibinfo {author} {\bibfnamefont {C.}~\bibnamefont
  {Zupancic}},\ }\href {\doibase 10.1103/PhysRev.129.854} {\bibfield  {journal}
  {\bibinfo  {journal} {Phys. Rev.}\ }\textbf {\bibinfo {volume} {129}},\
  \bibinfo {pages} {854} (\bibinfo {year} {1963})}\BibitemShut {NoStop}%
\bibitem [{\citenamefont {Sato}\ and\ \citenamefont
  {Yazaki}(1981)}]{Sato:1981ez}%
  \BibitemOpen
  \bibfield  {author} {\bibinfo {author} {\bibfnamefont {H.}~\bibnamefont
  {Sato}}\ and\ \bibinfo {author} {\bibfnamefont {K.}~\bibnamefont {Yazaki}},\
  }\href {\doibase 10.1016/0370-2693(81)90976-X} {\bibfield  {journal}
  {\bibinfo  {journal} {Phys. Lett. B}\ }\textbf {\bibinfo {volume} {98}},\
  \bibinfo {pages} {153} (\bibinfo {year} {1981})}\BibitemShut {NoStop}%
\bibitem [{\citenamefont {Mattiello}\ \emph {et~al.}(1995)\citenamefont
  {Mattiello}, \citenamefont {Jahns}, \citenamefont {Sorge}, \citenamefont
  {Stoecker},\ and\ \citenamefont {Greiner}}]{Mattiello:1995xg}%
  \BibitemOpen
  \bibfield  {author} {\bibinfo {author} {\bibfnamefont {R.}~\bibnamefont
  {Mattiello}}, \bibinfo {author} {\bibfnamefont {A.}~\bibnamefont {Jahns}},
  \bibinfo {author} {\bibfnamefont {H.}~\bibnamefont {Sorge}}, \bibinfo
  {author} {\bibfnamefont {H.}~\bibnamefont {Stoecker}}, \ and\ \bibinfo
  {author} {\bibfnamefont {W.}~\bibnamefont {Greiner}},\ }\href {\doibase
  10.1103/PhysRevLett.74.2180} {\bibfield  {journal} {\bibinfo  {journal}
  {Phys. Rev. Lett.}\ }\textbf {\bibinfo {volume} {74}},\ \bibinfo {pages}
  {2180} (\bibinfo {year} {1995})}\BibitemShut {NoStop}%
\bibitem [{\citenamefont {Nagle}\ \emph {et~al.}(1996)\citenamefont {Nagle},
  \citenamefont {Kumar}, \citenamefont {Kusnezov}, \citenamefont {Sorge},\ and\
  \citenamefont {Mattiello}}]{Nagle:1996vp}%
  \BibitemOpen
  \bibfield  {author} {\bibinfo {author} {\bibfnamefont {J.~L.}\ \bibnamefont
  {Nagle}}, \bibinfo {author} {\bibfnamefont {B.~S.}\ \bibnamefont {Kumar}},
  \bibinfo {author} {\bibfnamefont {D.}~\bibnamefont {Kusnezov}}, \bibinfo
  {author} {\bibfnamefont {H.}~\bibnamefont {Sorge}}, \ and\ \bibinfo {author}
  {\bibfnamefont {R.}~\bibnamefont {Mattiello}},\ }\href {\doibase
  10.1103/PhysRevC.53.367} {\bibfield  {journal} {\bibinfo  {journal} {Phys.
  Rev. C}\ }\textbf {\bibinfo {volume} {53}},\ \bibinfo {pages} {367} (\bibinfo
  {year} {1996})}\BibitemShut {NoStop}%
\bibitem [{\citenamefont {Mattiello}\ \emph {et~al.}(1997)\citenamefont
  {Mattiello}, \citenamefont {Sorge}, \citenamefont {Stoecker},\ and\
  \citenamefont {Greiner}}]{Mattiello:1996gq}%
  \BibitemOpen
  \bibfield  {author} {\bibinfo {author} {\bibfnamefont {R.}~\bibnamefont
  {Mattiello}}, \bibinfo {author} {\bibfnamefont {H.}~\bibnamefont {Sorge}},
  \bibinfo {author} {\bibfnamefont {H.}~\bibnamefont {Stoecker}}, \ and\
  \bibinfo {author} {\bibfnamefont {W.}~\bibnamefont {Greiner}},\ }\href
  {\doibase 10.1103/PhysRevC.55.1443} {\bibfield  {journal} {\bibinfo
  {journal} {Phys. Rev. C}\ }\textbf {\bibinfo {volume} {55}},\ \bibinfo
  {pages} {1443} (\bibinfo {year} {1997})},\ \Eprint
  {http://arxiv.org/abs/nucl-th/9607003} {arXiv:nucl-th/9607003} \BibitemShut
  {NoStop}%
\bibitem [{\citenamefont {Polleri}\ \emph {et~al.}(1998)\citenamefont
  {Polleri}, \citenamefont {Bondorf},\ and\ \citenamefont
  {Mishustin}}]{Polleri:1997bp}%
  \BibitemOpen
  \bibfield  {author} {\bibinfo {author} {\bibfnamefont {A.}~\bibnamefont
  {Polleri}}, \bibinfo {author} {\bibfnamefont {J.~P.}\ \bibnamefont
  {Bondorf}}, \ and\ \bibinfo {author} {\bibfnamefont {I.~N.}\ \bibnamefont
  {Mishustin}},\ }\href {\doibase 10.1016/S0370-2693(97)01455-X} {\bibfield
  {journal} {\bibinfo  {journal} {Phys. Lett. B}\ }\textbf {\bibinfo {volume}
  {419}},\ \bibinfo {pages} {19} (\bibinfo {year} {1998})},\ \Eprint
  {http://arxiv.org/abs/nucl-th/9711011} {arXiv:nucl-th/9711011} \BibitemShut
  {NoStop}%
\bibitem [{\citenamefont {Scheibl}\ and\ \citenamefont
  {Heinz}(1999)}]{Scheibl:1998tk}%
  \BibitemOpen
  \bibfield  {author} {\bibinfo {author} {\bibfnamefont {R.}~\bibnamefont
  {Scheibl}}\ and\ \bibinfo {author} {\bibfnamefont {U.~W.}\ \bibnamefont
  {Heinz}},\ }\href {\doibase 10.1103/PhysRevC.59.1585} {\bibfield  {journal}
  {\bibinfo  {journal} {Phys. Rev. C}\ }\textbf {\bibinfo {volume} {59}},\
  \bibinfo {pages} {1585} (\bibinfo {year} {1999})},\ \Eprint
  {http://arxiv.org/abs/nucl-th/9809092} {arXiv:nucl-th/9809092} \BibitemShut
  {NoStop}%
\bibitem [{\citenamefont {Sharma}\ \emph {et~al.}(2018)\citenamefont {Sharma},
  \citenamefont {Perez}, \citenamefont {Castro}, \citenamefont {Kumar},\ and\
  \citenamefont {Nattrass}}]{Sharma:2018dyb}%
  \BibitemOpen
  \bibfield  {author} {\bibinfo {author} {\bibfnamefont {N.}~\bibnamefont
  {Sharma}}, \bibinfo {author} {\bibfnamefont {T.}~\bibnamefont {Perez}},
  \bibinfo {author} {\bibfnamefont {A.}~\bibnamefont {Castro}}, \bibinfo
  {author} {\bibfnamefont {L.}~\bibnamefont {Kumar}}, \ and\ \bibinfo {author}
  {\bibfnamefont {C.}~\bibnamefont {Nattrass}},\ }\href {\doibase
  10.1103/PhysRevC.98.014914} {\bibfield  {journal} {\bibinfo  {journal} {Phys.
  Rev. C}\ }\textbf {\bibinfo {volume} {98}},\ \bibinfo {pages} {014914}
  (\bibinfo {year} {2018})},\ \Eprint {http://arxiv.org/abs/1803.02313}
  {arXiv:1803.02313 [hep-ph]} \BibitemShut {NoStop}%
\bibitem [{\citenamefont {Bazak}\ and\ \citenamefont
  {Mrowczynski}(2018)}]{Bazak:2018hgl}%
  \BibitemOpen
  \bibfield  {author} {\bibinfo {author} {\bibfnamefont {S.}~\bibnamefont
  {Bazak}}\ and\ \bibinfo {author} {\bibfnamefont {S.}~\bibnamefont
  {Mrowczynski}},\ }\href {\doibase 10.1142/S0217732318501420} {\bibfield
  {journal} {\bibinfo  {journal} {Mod. Phys. Lett. A}\ }\textbf {\bibinfo
  {volume} {33}},\ \bibinfo {pages} {1850142} (\bibinfo {year} {2018})},\
  \Eprint {http://arxiv.org/abs/1802.08212} {arXiv:1802.08212 [nucl-th]}
  \BibitemShut {NoStop}%
\bibitem [{\citenamefont {Danielewicz}\ and\ \citenamefont
  {Bertsch}(1991)}]{Danielewicz:1991dh}%
  \BibitemOpen
  \bibfield  {author} {\bibinfo {author} {\bibfnamefont {P.}~\bibnamefont
  {Danielewicz}}\ and\ \bibinfo {author} {\bibfnamefont {G.~F.}\ \bibnamefont
  {Bertsch}},\ }\href {\doibase 10.1016/0375-9474(91)90541-D} {\bibfield
  {journal} {\bibinfo  {journal} {Nucl. Phys. A}\ }\textbf {\bibinfo {volume}
  {533}},\ \bibinfo {pages} {712} (\bibinfo {year} {1991})}\BibitemShut
  {NoStop}%
\bibitem [{\citenamefont {Oh}\ \emph {et~al.}(2009)\citenamefont {Oh},
  \citenamefont {Lin},\ and\ \citenamefont {Ko}}]{Oh:2009gx}%
  \BibitemOpen
  \bibfield  {author} {\bibinfo {author} {\bibfnamefont {Y.}~\bibnamefont
  {Oh}}, \bibinfo {author} {\bibfnamefont {Z.-W.}\ \bibnamefont {Lin}}, \ and\
  \bibinfo {author} {\bibfnamefont {C.~M.}\ \bibnamefont {Ko}},\ }\href
  {\doibase 10.1103/PhysRevC.80.064902} {\bibfield  {journal} {\bibinfo
  {journal} {Phys. Rev. C}\ }\textbf {\bibinfo {volume} {80}},\ \bibinfo
  {pages} {064902} (\bibinfo {year} {2009})},\ \Eprint
  {http://arxiv.org/abs/0910.1977} {arXiv:0910.1977 [nucl-th]} \BibitemShut
  {NoStop}%
\bibitem [{\citenamefont {Oliinychenko}\ \emph
  {et~al.}(2019{\natexlab{a}})\citenamefont {Oliinychenko}, \citenamefont
  {Pang}, \citenamefont {Elfner},\ and\ \citenamefont
  {Koch}}]{Oliinychenko:2018ugs}%
  \BibitemOpen
  \bibfield  {author} {\bibinfo {author} {\bibfnamefont {D.}~\bibnamefont
  {Oliinychenko}}, \bibinfo {author} {\bibfnamefont {L.-G.}\ \bibnamefont
  {Pang}}, \bibinfo {author} {\bibfnamefont {H.}~\bibnamefont {Elfner}}, \ and\
  \bibinfo {author} {\bibfnamefont {V.}~\bibnamefont {Koch}},\ }\href {\doibase
  10.1103/PhysRevC.99.044907} {\bibfield  {journal} {\bibinfo  {journal} {Phys.
  Rev. C}\ }\textbf {\bibinfo {volume} {99}},\ \bibinfo {pages} {044907}
  (\bibinfo {year} {2019}{\natexlab{a}})},\ \Eprint
  {http://arxiv.org/abs/1809.03071} {arXiv:1809.03071 [hep-ph]} \BibitemShut
  {NoStop}%
\bibitem [{\citenamefont {Oliinychenko}\ \emph {et~al.}(2021)\citenamefont
  {Oliinychenko}, \citenamefont {Shen},\ and\ \citenamefont
  {Koch}}]{Oliinychenko:2020znl}%
  \BibitemOpen
  \bibfield  {author} {\bibinfo {author} {\bibfnamefont {D.}~\bibnamefont
  {Oliinychenko}}, \bibinfo {author} {\bibfnamefont {C.}~\bibnamefont {Shen}},
  \ and\ \bibinfo {author} {\bibfnamefont {V.}~\bibnamefont {Koch}},\ }\href
  {\doibase 10.1103/PhysRevC.103.034913} {\bibfield  {journal} {\bibinfo
  {journal} {Phys. Rev. C}\ }\textbf {\bibinfo {volume} {103}},\ \bibinfo
  {pages} {034913} (\bibinfo {year} {2021})},\ \Eprint
  {http://arxiv.org/abs/2009.01915} {arXiv:2009.01915 [hep-ph]} \BibitemShut
  {NoStop}%
\bibitem [{\citenamefont {Staudenmaier}\ \emph {et~al.}(2021)\citenamefont
  {Staudenmaier}, \citenamefont {Oliinychenko}, \citenamefont {Torres-Rincon},\
  and\ \citenamefont {Elfner}}]{Staudenmaier:2021lrg}%
  \BibitemOpen
  \bibfield  {author} {\bibinfo {author} {\bibfnamefont {J.}~\bibnamefont
  {Staudenmaier}}, \bibinfo {author} {\bibfnamefont {D.}~\bibnamefont
  {Oliinychenko}}, \bibinfo {author} {\bibfnamefont {J.~M.}\ \bibnamefont
  {Torres-Rincon}}, \ and\ \bibinfo {author} {\bibfnamefont {H.}~\bibnamefont
  {Elfner}},\ }\href {\doibase 10.1103/PhysRevC.104.034908} {\bibfield
  {journal} {\bibinfo  {journal} {Phys. Rev. C}\ }\textbf {\bibinfo {volume}
  {104}},\ \bibinfo {pages} {034908} (\bibinfo {year} {2021})},\ \Eprint
  {http://arxiv.org/abs/2106.14287} {arXiv:2106.14287 [hep-ph]} \BibitemShut
  {NoStop}%
\bibitem [{\citenamefont {Kireyeu}\ \emph {et~al.}(2022)\citenamefont
  {Kireyeu}, \citenamefont {Steinheimer}, \citenamefont {Aichelin},
  \citenamefont {Bleicher},\ and\ \citenamefont
  {Bratkovskaya}}]{Kireyeu:2022qmv}%
  \BibitemOpen
  \bibfield  {author} {\bibinfo {author} {\bibfnamefont {V.}~\bibnamefont
  {Kireyeu}}, \bibinfo {author} {\bibfnamefont {J.}~\bibnamefont
  {Steinheimer}}, \bibinfo {author} {\bibfnamefont {J.}~\bibnamefont
  {Aichelin}}, \bibinfo {author} {\bibfnamefont {M.}~\bibnamefont {Bleicher}},
  \ and\ \bibinfo {author} {\bibfnamefont {E.}~\bibnamefont {Bratkovskaya}},\
  }\href {\doibase 10.1103/PhysRevC.105.044909} {\bibfield  {journal} {\bibinfo
   {journal} {Phys. Rev. C}\ }\textbf {\bibinfo {volume} {105}},\ \bibinfo
  {pages} {044909} (\bibinfo {year} {2022})},\ \Eprint
  {http://arxiv.org/abs/2201.13374} {arXiv:2201.13374 [nucl-th]} \BibitemShut
  {NoStop}%
\bibitem [{\citenamefont {Coci}\ \emph {et~al.}(2023)\citenamefont {Coci},
  \citenamefont {Gl\"a\ss{}el}, \citenamefont {Kireyeu}, \citenamefont
  {Aichelin}, \citenamefont {Blume}, \citenamefont {Bratkovskaya},
  \citenamefont {Kolesnikov},\ and\ \citenamefont {Voronyuk}}]{Coci:2023daq}%
  \BibitemOpen
  \bibfield  {author} {\bibinfo {author} {\bibfnamefont {G.}~\bibnamefont
  {Coci}}, \bibinfo {author} {\bibfnamefont {S.}~\bibnamefont {Gl\"a\ss{}el}},
  \bibinfo {author} {\bibfnamefont {V.}~\bibnamefont {Kireyeu}}, \bibinfo
  {author} {\bibfnamefont {J.}~\bibnamefont {Aichelin}}, \bibinfo {author}
  {\bibfnamefont {C.}~\bibnamefont {Blume}}, \bibinfo {author} {\bibfnamefont
  {E.}~\bibnamefont {Bratkovskaya}}, \bibinfo {author} {\bibfnamefont
  {V.}~\bibnamefont {Kolesnikov}}, \ and\ \bibinfo {author} {\bibfnamefont
  {V.}~\bibnamefont {Voronyuk}},\ }\href {\doibase 10.1103/PhysRevC.108.014902}
  {\bibfield  {journal} {\bibinfo  {journal} {Phys. Rev. C}\ }\textbf {\bibinfo
  {volume} {108}},\ \bibinfo {pages} {014902} (\bibinfo {year} {2023})},\
  \Eprint {http://arxiv.org/abs/2303.02279} {arXiv:2303.02279 [nucl-th]}
  \BibitemShut {NoStop}%
\bibitem [{\citenamefont {Zhao}\ \emph {et~al.}(2022)\citenamefont {Zhao},
  \citenamefont {Feng}, \citenamefont {Shao}, \citenamefont {Wang},\ and\
  \citenamefont {Song}}]{Zhao:2022xkz}%
  \BibitemOpen
  \bibfield  {author} {\bibinfo {author} {\bibfnamefont {X.-Y.}\ \bibnamefont
  {Zhao}}, \bibinfo {author} {\bibfnamefont {Y.-T.}\ \bibnamefont {Feng}},
  \bibinfo {author} {\bibfnamefont {F.-L.}\ \bibnamefont {Shao}}, \bibinfo
  {author} {\bibfnamefont {R.-Q.}\ \bibnamefont {Wang}}, \ and\ \bibinfo
  {author} {\bibfnamefont {J.}~\bibnamefont {Song}},\ }\href {\doibase
  10.1103/PhysRevC.105.054908} {\bibfield  {journal} {\bibinfo  {journal}
  {Phys. Rev. C}\ }\textbf {\bibinfo {volume} {105}},\ \bibinfo {pages}
  {054908} (\bibinfo {year} {2022})},\ \Eprint
  {http://arxiv.org/abs/2201.10354} {arXiv:2201.10354 [hep-ph]} \BibitemShut
  {NoStop}%
\bibitem [{\citenamefont {Wang}\ \emph {et~al.}(2022)\citenamefont {Wang},
  \citenamefont {Lv}, \citenamefont {Li}, \citenamefont {Song},\ and\
  \citenamefont {Shao}}]{Wang:2022hja}%
  \BibitemOpen
  \bibfield  {author} {\bibinfo {author} {\bibfnamefont {R.-Q.}\ \bibnamefont
  {Wang}}, \bibinfo {author} {\bibfnamefont {J.-P.}\ \bibnamefont {Lv}},
  \bibinfo {author} {\bibfnamefont {Y.-H.}\ \bibnamefont {Li}}, \bibinfo
  {author} {\bibfnamefont {J.}~\bibnamefont {Song}}, \ and\ \bibinfo {author}
  {\bibfnamefont {F.-L.}\ \bibnamefont {Shao}},\ }\href@noop {} {\  (\bibinfo
  {year} {2022})},\ \Eprint {http://arxiv.org/abs/2210.10271} {arXiv:2210.10271
  [hep-ph]} \BibitemShut {NoStop}%
\bibitem [{\citenamefont {Wang}\ \emph {et~al.}(2021)\citenamefont {Wang},
  \citenamefont {Shao},\ and\ \citenamefont {Song}}]{Wang:2020zaw}%
  \BibitemOpen
  \bibfield  {author} {\bibinfo {author} {\bibfnamefont {R.-Q.}\ \bibnamefont
  {Wang}}, \bibinfo {author} {\bibfnamefont {F.-L.}\ \bibnamefont {Shao}}, \
  and\ \bibinfo {author} {\bibfnamefont {J.}~\bibnamefont {Song}},\ }\href
  {\doibase 10.1103/PhysRevC.103.064908} {\bibfield  {journal} {\bibinfo
  {journal} {Phys. Rev. C}\ }\textbf {\bibinfo {volume} {103}},\ \bibinfo
  {pages} {064908} (\bibinfo {year} {2021})},\ \Eprint
  {http://arxiv.org/abs/2007.05745} {arXiv:2007.05745 [hep-ph]} \BibitemShut
  {NoStop}%
\bibitem [{\citenamefont {Csorgo}\ and\ \citenamefont
  {Lorstad}(1996)}]{Csorgo:1995bi}%
  \BibitemOpen
  \bibfield  {author} {\bibinfo {author} {\bibfnamefont {T.}~\bibnamefont
  {Csorgo}}\ and\ \bibinfo {author} {\bibfnamefont {B.}~\bibnamefont
  {Lorstad}},\ }\href {\doibase 10.1103/PhysRevC.54.1390} {\bibfield  {journal}
  {\bibinfo  {journal} {Phys. Rev. C}\ }\textbf {\bibinfo {volume} {54}},\
  \bibinfo {pages} {1390} (\bibinfo {year} {1996})},\ \Eprint
  {http://arxiv.org/abs/hep-ph/9509213} {arXiv:hep-ph/9509213} \BibitemShut
  {NoStop}%
\bibitem [{\citenamefont {Tomasik}\ and\ \citenamefont
  {Heinz}(1998)}]{Tomasik:1997eq}%
  \BibitemOpen
  \bibfield  {author} {\bibinfo {author} {\bibfnamefont {B.}~\bibnamefont
  {Tomasik}}\ and\ \bibinfo {author} {\bibfnamefont {U.~W.}\ \bibnamefont
  {Heinz}},\ }\href {\doibase 10.1007/s100520050211} {\bibfield  {journal}
  {\bibinfo  {journal} {Eur. Phys. J. C}\ }\textbf {\bibinfo {volume} {4}},\
  \bibinfo {pages} {327} (\bibinfo {year} {1998})},\ \Eprint
  {http://arxiv.org/abs/nucl-th/9707001} {arXiv:nucl-th/9707001} \BibitemShut
  {NoStop}%
\bibitem [{\citenamefont {Wang}\ \emph {et~al.}(2019)\citenamefont {Wang},
  \citenamefont {Song}, \citenamefont {Li},\ and\ \citenamefont
  {Shao}}]{Wang:2017vsm}%
  \BibitemOpen
  \bibfield  {author} {\bibinfo {author} {\bibfnamefont {R.-q.}\ \bibnamefont
  {Wang}}, \bibinfo {author} {\bibfnamefont {J.}~\bibnamefont {Song}}, \bibinfo
  {author} {\bibfnamefont {G.}~\bibnamefont {Li}}, \ and\ \bibinfo {author}
  {\bibfnamefont {F.-l.}\ \bibnamefont {Shao}},\ }\href {\doibase
  10.1088/1674-1137/43/2/024101} {\bibfield  {journal} {\bibinfo  {journal}
  {Chin. Phys. C}\ }\textbf {\bibinfo {volume} {43}},\ \bibinfo {pages}
  {024101} (\bibinfo {year} {2019})},\ \Eprint
  {http://arxiv.org/abs/1710.08572} {arXiv:1710.08572 [hep-ph]} \BibitemShut
  {NoStop}%
\bibitem [{\citenamefont {Chen}\ \emph
  {et~al.}(2003{\natexlab{b}})\citenamefont {Chen}, \citenamefont {Ko},\ and\
  \citenamefont {Li}}]{Chen:2003ava}%
  \BibitemOpen
  \bibfield  {author} {\bibinfo {author} {\bibfnamefont {L.-W.}\ \bibnamefont
  {Chen}}, \bibinfo {author} {\bibfnamefont {C.~M.}\ \bibnamefont {Ko}}, \ and\
  \bibinfo {author} {\bibfnamefont {B.-A.}\ \bibnamefont {Li}},\ }\href
  {\doibase 10.1016/j.nuclphysa.2003.09.010} {\bibfield  {journal} {\bibinfo
  {journal} {Nucl. Phys. A}\ }\textbf {\bibinfo {volume} {729}},\ \bibinfo
  {pages} {809} (\bibinfo {year} {2003}{\natexlab{b}})},\ \Eprint
  {http://arxiv.org/abs/nucl-th/0306032} {arXiv:nucl-th/0306032} \BibitemShut
  {NoStop}%
\bibitem [{\citenamefont {Zhu}\ \emph {et~al.}(2015)\citenamefont {Zhu},
  \citenamefont {Ko},\ and\ \citenamefont {Yin}}]{Zhu:2015voa}%
  \BibitemOpen
  \bibfield  {author} {\bibinfo {author} {\bibfnamefont {L.}~\bibnamefont
  {Zhu}}, \bibinfo {author} {\bibfnamefont {C.~M.}\ \bibnamefont {Ko}}, \ and\
  \bibinfo {author} {\bibfnamefont {X.}~\bibnamefont {Yin}},\ }\href {\doibase
  10.1103/PhysRevC.92.064911} {\bibfield  {journal} {\bibinfo  {journal} {Phys.
  Rev. C}\ }\textbf {\bibinfo {volume} {92}},\ \bibinfo {pages} {064911}
  (\bibinfo {year} {2015})},\ \Eprint {http://arxiv.org/abs/1510.03568}
  {arXiv:1510.03568 [nucl-th]} \BibitemShut {NoStop}%
\bibitem [{\citenamefont {Angeli}\ and\ \citenamefont
  {Marinova}(2013)}]{Angeli:2013epw}%
  \BibitemOpen
  \bibfield  {author} {\bibinfo {author} {\bibfnamefont {I.}~\bibnamefont
  {Angeli}}\ and\ \bibinfo {author} {\bibfnamefont {K.~P.}\ \bibnamefont
  {Marinova}},\ }\href {\doibase 10.1016/j.adt.2011.12.006} {\bibfield
  {journal} {\bibinfo  {journal} {Atom. Data Nucl. Data Tabl.}\ }\textbf
  {\bibinfo {volume} {99}},\ \bibinfo {pages} {69} (\bibinfo {year}
  {2013})}\BibitemShut {NoStop}%
\bibitem [{\citenamefont {Mrowczynski}(2017)}]{Mrowczynski:2016xqm}%
  \BibitemOpen
  \bibfield  {author} {\bibinfo {author} {\bibfnamefont {S.}~\bibnamefont
  {Mrowczynski}},\ }\href {\doibase 10.5506/APhysPolB.48.707} {\bibfield
  {journal} {\bibinfo  {journal} {Acta Phys. Polon. B}\ }\textbf {\bibinfo
  {volume} {48}},\ \bibinfo {pages} {707} (\bibinfo {year} {2017})},\ \Eprint
  {http://arxiv.org/abs/1607.02267} {arXiv:1607.02267 [nucl-th]} \BibitemShut
  {NoStop}%
\bibitem [{\citenamefont {Kisiel}\ \emph {et~al.}(2014)\citenamefont {Kisiel},
  \citenamefont {Ga\l{}a\.zyn},\ and\ \citenamefont
  {Bo\.zek}}]{Kisiel:2014upa}%
  \BibitemOpen
  \bibfield  {author} {\bibinfo {author} {\bibfnamefont {A.}~\bibnamefont
  {Kisiel}}, \bibinfo {author} {\bibfnamefont {M.}~\bibnamefont
  {Ga\l{}a\.zyn}}, \ and\ \bibinfo {author} {\bibfnamefont {P.}~\bibnamefont
  {Bo\.zek}},\ }\href {\doibase 10.1103/PhysRevC.90.064914} {\bibfield
  {journal} {\bibinfo  {journal} {Phys. Rev. C}\ }\textbf {\bibinfo {volume}
  {90}},\ \bibinfo {pages} {064914} (\bibinfo {year} {2014})},\ \Eprint
  {http://arxiv.org/abs/1409.4571} {arXiv:1409.4571 [nucl-th]} \BibitemShut
  {NoStop}%
\bibitem [{\citenamefont {Adam}\ \emph {et~al.}(2015)\citenamefont {Adam} \emph
  {et~al.}}]{ALICE:2015hvw}%
  \BibitemOpen
  \bibfield  {author} {\bibinfo {author} {\bibfnamefont {J.}~\bibnamefont
  {Adam}} \emph {et~al.} (\bibinfo {collaboration} {ALICE}),\ }\href {\doibase
  10.1103/PhysRevC.92.054908} {\bibfield  {journal} {\bibinfo  {journal} {Phys.
  Rev. C}\ }\textbf {\bibinfo {volume} {92}},\ \bibinfo {pages} {054908}
  (\bibinfo {year} {2015})},\ \Eprint {http://arxiv.org/abs/1506.07884}
  {arXiv:1506.07884 [nucl-ex]} \BibitemShut {NoStop}%
\bibitem [{\citenamefont {Adam}\ \emph
  {et~al.}(2016{\natexlab{b}})\citenamefont {Adam} \emph
  {et~al.}}]{ALICE:2015tra}%
  \BibitemOpen
  \bibfield  {author} {\bibinfo {author} {\bibfnamefont {J.}~\bibnamefont
  {Adam}} \emph {et~al.} (\bibinfo {collaboration} {ALICE}),\ }\href {\doibase
  10.1103/PhysRevC.93.024905} {\bibfield  {journal} {\bibinfo  {journal} {Phys.
  Rev. C}\ }\textbf {\bibinfo {volume} {93}},\ \bibinfo {pages} {024905}
  (\bibinfo {year} {2016}{\natexlab{b}})},\ \Eprint
  {http://arxiv.org/abs/1507.06842} {arXiv:1507.06842 [nucl-ex]} \BibitemShut
  {NoStop}%
\bibitem [{\citenamefont {Acharya}\ \emph {et~al.}(2017)\citenamefont {Acharya}
  \emph {et~al.}}]{ALICE:2017nuf}%
  \BibitemOpen
  \bibfield  {author} {\bibinfo {author} {\bibfnamefont {S.}~\bibnamefont
  {Acharya}} \emph {et~al.} (\bibinfo {collaboration} {ALICE}),\ }\href
  {\doibase 10.1140/epjc/s10052-017-5222-x} {\bibfield  {journal} {\bibinfo
  {journal} {Eur. Phys. J. C}\ }\textbf {\bibinfo {volume} {77}},\ \bibinfo
  {pages} {658} (\bibinfo {year} {2017})},\ \Eprint
  {http://arxiv.org/abs/1707.07304} {arXiv:1707.07304 [nucl-ex]} \BibitemShut
  {NoStop}%
\bibitem [{\citenamefont {Chakraborty}\ \emph {et~al.}(2021)\citenamefont
  {Chakraborty}, \citenamefont {Pandey},\ and\ \citenamefont
  {Dash}}]{Chakraborty:2020tym}%
  \BibitemOpen
  \bibfield  {author} {\bibinfo {author} {\bibfnamefont {P.}~\bibnamefont
  {Chakraborty}}, \bibinfo {author} {\bibfnamefont {A.~K.}\ \bibnamefont
  {Pandey}}, \ and\ \bibinfo {author} {\bibfnamefont {S.}~\bibnamefont
  {Dash}},\ }\href {\doibase 10.1140/epja/s10050-021-00647-w} {\bibfield
  {journal} {\bibinfo  {journal} {Eur. Phys. J. A}\ }\textbf {\bibinfo {volume}
  {57}},\ \bibinfo {pages} {338} (\bibinfo {year} {2021})},\ \Eprint
  {http://arxiv.org/abs/2010.12161} {arXiv:2010.12161 [hep-ph]} \BibitemShut
  {NoStop}%
\bibitem [{\citenamefont {Adams}\ \emph {et~al.}(2005)\citenamefont {Adams}
  \emph {et~al.}}]{STAR:2004qya}%
  \BibitemOpen
  \bibfield  {author} {\bibinfo {author} {\bibfnamefont {J.}~\bibnamefont
  {Adams}} \emph {et~al.} (\bibinfo {collaboration} {STAR}),\ }\href {\doibase
  10.1103/PhysRevC.71.044906} {\bibfield  {journal} {\bibinfo  {journal} {Phys.
  Rev. C}\ }\textbf {\bibinfo {volume} {71}},\ \bibinfo {pages} {044906}
  (\bibinfo {year} {2005})},\ \Eprint {http://arxiv.org/abs/nucl-ex/0411036}
  {arXiv:nucl-ex/0411036} \BibitemShut {NoStop}%
\bibitem [{\citenamefont {Abelev}\ \emph {et~al.}(2013)\citenamefont {Abelev}
  \emph {et~al.}}]{ALICE:2013mez}%
  \BibitemOpen
  \bibfield  {author} {\bibinfo {author} {\bibfnamefont {B.}~\bibnamefont
  {Abelev}} \emph {et~al.} (\bibinfo {collaboration} {ALICE}),\ }\href
  {\doibase 10.1103/PhysRevC.88.044910} {\bibfield  {journal} {\bibinfo
  {journal} {Phys. Rev. C}\ }\textbf {\bibinfo {volume} {88}},\ \bibinfo
  {pages} {044910} (\bibinfo {year} {2013})},\ \Eprint
  {http://arxiv.org/abs/1303.0737} {arXiv:1303.0737 [hep-ex]} \BibitemShut
  {NoStop}%
\bibitem [{\citenamefont {Oliinychenko}\ \emph
  {et~al.}(2019{\natexlab{b}})\citenamefont {Oliinychenko}, \citenamefont
  {Pang}, \citenamefont {Elfner},\ and\ \citenamefont
  {Koch}}]{Oliinychenko:2018odl}%
  \BibitemOpen
  \bibfield  {author} {\bibinfo {author} {\bibfnamefont {D.}~\bibnamefont
  {Oliinychenko}}, \bibinfo {author} {\bibfnamefont {L.-G.}\ \bibnamefont
  {Pang}}, \bibinfo {author} {\bibfnamefont {H.}~\bibnamefont {Elfner}}, \ and\
  \bibinfo {author} {\bibfnamefont {V.}~\bibnamefont {Koch}},\ }\href {\doibase
  10.3390/proceedings2019010006} {\bibfield  {journal} {\bibinfo  {journal}
  {MDPI Proc.}\ }\textbf {\bibinfo {volume} {10}},\ \bibinfo {pages} {6}
  (\bibinfo {year} {2019}{\natexlab{b}})},\ \Eprint
  {http://arxiv.org/abs/1812.06225} {arXiv:1812.06225 [hep-ph]} \BibitemShut
  {NoStop}%
\bibitem [{\citenamefont {Bailung}\ \emph {et~al.}(2023)\citenamefont
  {Bailung}, \citenamefont {Shah},\ and\ \citenamefont
  {Roy}}]{Bailung:2023dpv}%
  \BibitemOpen
  \bibfield  {author} {\bibinfo {author} {\bibfnamefont {Y.}~\bibnamefont
  {Bailung}}, \bibinfo {author} {\bibfnamefont {N.}~\bibnamefont {Shah}}, \
  and\ \bibinfo {author} {\bibfnamefont {A.}~\bibnamefont {Roy}},\ }\href
  {\doibase 10.1016/j.nuclphysa.2023.122701} {\bibfield  {journal} {\bibinfo
  {journal} {Nucl. Phys. A}\ }\textbf {\bibinfo {volume} {1037}},\ \bibinfo
  {pages} {122701} (\bibinfo {year} {2023})}\BibitemShut {NoStop}%
\bibitem [{\citenamefont {Liu}\ \emph {et~al.}(2022)\citenamefont {Liu},
  \citenamefont {She}, \citenamefont {Xu}, \citenamefont {Zhou}, \citenamefont
  {Chen},\ and\ \citenamefont {Sa}}]{Liu:2022vbg}%
  \BibitemOpen
  \bibfield  {author} {\bibinfo {author} {\bibfnamefont {F.-X.}\ \bibnamefont
  {Liu}}, \bibinfo {author} {\bibfnamefont {Z.-L.}\ \bibnamefont {She}},
  \bibinfo {author} {\bibfnamefont {H.-G.}\ \bibnamefont {Xu}}, \bibinfo
  {author} {\bibfnamefont {D.-M.}\ \bibnamefont {Zhou}}, \bibinfo {author}
  {\bibfnamefont {G.}~\bibnamefont {Chen}}, \ and\ \bibinfo {author}
  {\bibfnamefont {B.-H.}\ \bibnamefont {Sa}},\ }\href {\doibase
  10.1038/s41598-022-05584-2} {\bibfield  {journal} {\bibinfo  {journal} {Sci.
  Rep.}\ }\textbf {\bibinfo {volume} {12}},\ \bibinfo {pages} {1772} (\bibinfo
  {year} {2022})}\BibitemShut {NoStop}%
\bibitem [{\citenamefont {Schnedermann}\ \emph {et~al.}(1993)\citenamefont
  {Schnedermann}, \citenamefont {Sollfrank},\ and\ \citenamefont
  {Heinz}}]{Schnedermann:1993ws}%
  \BibitemOpen
  \bibfield  {author} {\bibinfo {author} {\bibfnamefont {E.}~\bibnamefont
  {Schnedermann}}, \bibinfo {author} {\bibfnamefont {J.}~\bibnamefont
  {Sollfrank}}, \ and\ \bibinfo {author} {\bibfnamefont {U.~W.}\ \bibnamefont
  {Heinz}},\ }\href {\doibase 10.1103/PhysRevC.48.2462} {\bibfield  {journal}
  {\bibinfo  {journal} {Phys. Rev. C}\ }\textbf {\bibinfo {volume} {48}},\
  \bibinfo {pages} {2462} (\bibinfo {year} {1993})},\ \Eprint
  {http://arxiv.org/abs/nucl-th/9307020} {arXiv:nucl-th/9307020} \BibitemShut
  {NoStop}%
\bibitem [{\citenamefont {Reichert}\ \emph {et~al.}(2023)\citenamefont
  {Reichert}, \citenamefont {Steinheimer}, \citenamefont {Vovchenko},
  \citenamefont {D\"onigus},\ and\ \citenamefont
  {Bleicher}}]{Reichert:2022mek}%
  \BibitemOpen
  \bibfield  {author} {\bibinfo {author} {\bibfnamefont {T.}~\bibnamefont
  {Reichert}}, \bibinfo {author} {\bibfnamefont {J.}~\bibnamefont
  {Steinheimer}}, \bibinfo {author} {\bibfnamefont {V.}~\bibnamefont
  {Vovchenko}}, \bibinfo {author} {\bibfnamefont {B.}~\bibnamefont
  {D\"onigus}}, \ and\ \bibinfo {author} {\bibfnamefont {M.}~\bibnamefont
  {Bleicher}},\ }\href {\doibase 10.1103/PhysRevC.107.014912} {\bibfield
  {journal} {\bibinfo  {journal} {Phys. Rev. C}\ }\textbf {\bibinfo {volume}
  {107}},\ \bibinfo {pages} {014912} (\bibinfo {year} {2023})},\ \Eprint
  {http://arxiv.org/abs/2210.11876} {arXiv:2210.11876 [nucl-th]} \BibitemShut
  {NoStop}%
\end{thebibliography}%

\end{document}